\documentclass{emulateapj}
\usepackage{apjfonts}
\usepackage{natbib}
\usepackage{amsfonts}
\usepackage{amsmath}
\usepackage[usenames]{color}
\usepackage{verbatim}
\newcommand{\jfm}{\rmfamily J.~Fluid~Mech.}%
\newcommand{\arfm}{\rmfamily Ann.~Review~Fluid~Mech.}%

\def\sun{_\odot}

\def\msun{M$_\odot$}

\def\Dwa{$\,$\uppercase\expandafter{\romannumeral5}$\,$}

\def\sless{\lower2pt\hbox{$\buildrel {\scriptstyle <}
   \over {\scriptstyle\sim}$}}

\def\sgreat{\lower2pt\hbox{$\buildrel {\scriptstyle >}
   \over {\scriptstyle\sim}$}}
\def\sharpnull#1{}

\newcommand{\Ibar}{I\hspace{-.5em}{\scriptscriptstyle \stackrel{\mathbf{--}}{}}}

\begin{document}




\title{A Model for Gravitational Wave Emission from Neutrino-Driven
  Core-Collapse Supernovae}


\author{Jeremiah W. Murphy\altaffilmark{1,2}}
\author{Christian D. Ott\altaffilmark{3,4,5}}
\author{Adam Burrows\altaffilmark{6}}

\altaffiltext{1}{Department of Astronomy, University of Washington,
  Box 351580, Seattle, WA 98195-1580, USA; jmurphy@astro.washington.edu}
\altaffiltext{2}{NSF Astronomy and Astrophysics Postdoctoral Fellow}
\altaffiltext{3}{Theoretical Astrophysics, Mailcode 350-17, 
  California Institute of Technology, Pasadena, CA 91125, USA;
  cott@tapir.caltech.edu}
\altaffiltext{4}{Niels Bohr International Academy, Niels Bohr Institute, 
  Blegdamsvej 17, DK-2100 Copenhagen, Denmark}
\altaffiltext{5}{Center for Computation and Technology, Louisiana State University, 
  Baton Rogue, LA, USA}
\altaffiltext{6}{Department of Astrophysical Sciences,
  Princeton University, Peyton Hall, Ivy Lane,
  Princeton, NJ 08544,USA; burrows@astro.princeton.edu} 

\begin{abstract}
Using a suite of progenitor models, neutrino luminosities, and
  two-dimensional (2D) simulations, we investigate the matter
  gravitational-wave (GW) emission from postbounce phases of
  neutrino-driven core-collapse supernovae (CCSNe).  These phases
  include prompt and steady-state convection, the standing accretion
  shock instability (SASI), and asymmetric explosions. For the stages
  before explosion, we propose a model for the source of GW emission.
  Downdrafts of the postshock-convection/SASI region strike the
  protoneutron star ``surface'' with large speeds and are decelerated
  by buoyancy forces.  We find that the GW amplitude is set by the
  magnitude of deceleration and, by extension, the downdraft's speed
  and the vigor of postshock-convective/SASI motions. However, the
  characteristic frequencies, which evolve from $\sim$100 Hz after
  bounce to $\sim$300-400 Hz, are practically independent of these
  speeds (and turnover timescales). Instead, they are set by the deceleration
  timescale, which is in turn set by the buoyancy frequency at the
  lower boundary of postshock convection. Consequently, the 
  characteristic GW frequencies are dependent upon a combination of core
  structure attributes, specifically the dense-matter equation of
  state (EOS) and details that determine the gradients at the
  boundary, including the accretion-rate history, the EOS at
  subnuclear densities, and neutrino transport. During explosion, the
  high frequency signal wanes and is replaced by a strong low
  frequency, $\sim$10s of Hz, signal that reveals the general
  morphology of the explosion (i.e. prolate, oblate, or
  spherical). However, current and near-future GW detectors are
  sensitive to GW power at frequencies $\gtrsim$50 Hz.  Therefore, the
  signature of explosion will be the abrupt reduction of detectable GW
  emission.

\end{abstract}


\section{Introduction}
\label{section:intro}

Core-collapse supernovae (CCSNe) are among the most energetic events in the
Universe; they herald the birth of neutron stars and black holes, are
a major site for nucleosynthesis, influence galactic dynamics,
trigger further star formation, and are prodigious emitters of
neutrinos and gravitational waves.  Hence, it is important to
understand the mechanism of explosion, yet the details have remained
elusive for many decades.

Gravitational waves (GWs) promise to be important
diagnostics for revealing the secrets of the CCSN mechanism.  Despite
the elusiveness of the CCSN problem, current multi-dimensional
simulations show several promising mechanisms.  Whether the solution
is the convection/SASI-aided neutrino mechanism \citep{marek09b,murphy08b}, the
acoustic mechanism \citep{burrows:06,burrows:07a}, or
magnetohydrodynamic (MHD) jets (\citealt{burrows:07b,dessart:08a} and
references therein), viable explosion mechanisms for all but the least
massive of massive stars \citep{kitaura:06,burrows:07c} appear
fundamentally multi-dimensional.  Because photons couple strongly with
matter, the mechanism is obscured by many solar masses and traditional
means of observation are limited.  However, neutrinos and GWs interact very
weakly with matter and provide a clear view into the dynamics of the
mechanism.  Unlike neutrinos, GWs are emitted (at lowest
order) by time-changing quadrupole mass/energy motions.  Therefore, the
multi-dimensional mechanisms for CCSNe are expected to be a
significant source of GWs, which will in turn provide a diagnostic
into their multi-dimensional nature.

This important connection between CCSNe and GWs was recognized quite
early \citep[for a comprehensive review of GWs and CCSNe, see][]
{ott:09a}. Most early investigations were limited to the collapse and
bounce of rapidly rotating iron cores \citep{mueller:82,moenchmeyer:91,yamadasato:95,zwerger:97,dimmelmeier:02,kotake:03,ott:04}.
Current investigations, that focus on these early phases and rapid
rotation, use 2D or 3D simulations, conformally-flat or full general
relativity, finite-temperature nuclear equations of state (EOS), and
deleptonization during collapse
\citep{dimmelmeier:07,dimmelmeier:08,ott:07prl}. However, the rotation
rates necessary for GWs with sizeable amplitudes are likely to
obtain in only $\sim$1\% of the massive star population
\citep{heger:05,hirschi04,woosley06,ott:06spin}.

Another set of studies has focused on the GW signatures of
postbounce phases and dynamics.  These include
convection in the protoneutron star (PNS) and in the postshock region
\citep{burrows96,mueller97,mueller:04,marek09a}, the standing
accretion shock instability (SASI) \citep{kotake07,kotake09,marek09a},
PNS internal g-modes associated with an acoustic mechanism for
explosion \citep{burrows:06,burrows:07a,ott:06b}, rapid rotation in
the context of MHD jets \citep{kotake:04,obergaulinger:06a}, or
asymmetric explosion and neutrino emission which lead to a long-term
nonzero GW strain, or ``memory''
\citep{braginskii87,burrows96,mueller97,mueller:04}.  Each signature
is associated with the three possible explosion mechanisms, and
\citet{ott:09b} suggests that their GW characteristics may help to
distinguish which mechanism obtains in a particular supernova.

In this paper, we do not address the GW signatures of PNS g-modes nor
rapid rotation but focus on the GWs from the postbounce phases of the
neutrino mechanism.  The bounce shock is quickly stalled by nuclear
dissociation, electron capture, and neutrino losses
\citep{mazurek82,bruenn:85,bruenn89} and leaves a negative entropy
gradient.  This leads to an initial, short-lived ($\lesssim 50$ ms)
phase of strong convective overturn, prompt convection \citep{burrows92}, and an
associated burst of GWs \citep{marek09a,ott:09a}.  More than two decades
ago, \citet{wilson:85} and \citet{bethewilson:85} suggested that a
fraction of the neutrinos being emitted from depth ($\lesssim$100 km)
would be recaptured in
the gain region ($\gtrsim$100 km), reviving the stalled shock into explosion.  However,
detailed one-dimensional (1D) simulations have shown that this
mechanism, the neutrino mechanism, fails in 1D
\citep{liebendoerfer:01a, liebendoerfer:01b, ramppjanka:02, buras:03,
  thompson:03, liebendoerfer:05}, except for the least massive of the
massive stars \citep{kitaura:06,burrows:07c}.  Recent 2D simulations
that are subject to aspherical instabilities, specifically
postshock convection, which is driven by neutrino heating in the gain
region and the SASI, suggest that the neutrino
mechanism may yet succeed, though it fails in 1D \citep{herant:94,
  janka95, bhf:95, jankamueller:96, burrows:07a, kitaura:06,
  buras:06a, buras:06b, marek09b, ott:08, murphy08b}.  In particular,
\citet{murphy08b} parametrized the neutrino luminosity and found that
the critical neutrino luminosity for explosions is lower in 2D
simulations compared to 1D.  These results suggest that asymmetries
are not only important for the production of GWs, but also for the
success of the neutrino mechanism.

PNS convection (see \citet{dessart:06a} and the references therein)
and postshock convection
\citep{bhf:95,jankamueller:96} have been recognized as an important
ingredient in CCSNe for many decades.  However, it was only recently
that the SASI has been recognized as a distinct instability that
augments neutrino-driven convection in the gain region \citep{blondin:03}.  The
mechanism responsible for this instability is debated to be either an
advective-acoustic cycle
\citep{foglizzo:00,foglizzo:01,foglizzo:02,blondin:03,ohnishi06,foglizzo:07,iwakami:08,scheck:08,yamasaki:08,foglizzo09,sato09}
or purely acoustic cycle \citep{blondin:06,laming07}, though recent
linear and nonlinear analysis of simple models suggest that the
advective-acoustic cycle is responsible \citep{sato09,foglizzo09}.
An additional complication is that it is hard to
disentangle the effects of the SASI from those of postshock
convection.
They both occupy the same region, reach nonlinear
saturation quickly, and likely influence each other's nonlinear
motions.  Given the recent recognition of the SASI, it is no wonder that few
have investigated fully the GW characteristics of the convection/SASI-aided
neutrino mechanism \citep{kotake07,kotake09,marek09a}.

Using simulations performed with the 2D radiation-hydrodynamics code
VULCAN/2D, \citet{ott:06b} described some GW signatures of the SASI,
but primarily focused on the GW emission from strong core g-modes.
\citet{kotake07} were the first to specifically investigate the GW
signature of the SASI in 2D.  To expedite calculations, they made
several assumptions.  Their initial conditions were derived from a
steady-state approximation of the postshock structure and not from
stellar evolution calculations.  They used an inner boundary at 50 km
and kept the density at this boundary constant at $10^{11}$ g
cm$^{-3}$.  Although they used a relativistic-mean-field 
EOS \citep{shen:98a}, they approximated the effects of
neutrino interactions with local heating and cooling terms and assumed
a fixed accretion rate at the grid outer boundary.  Even
though they omitted the PNS and proper neutrino transport, they
estimated the GW emission from matter and neutrinos, and found that
the GW amplitudes due to asymmetric neutrino emission are $\sim$100
times that of matter, but at low frequency and hence lower power.
They also found that the matter GW signal grows as the SASI enters the
nonlinear phase and that the characteristic frequency from the matter
GW emission is $\sim$100 Hz.  Interestingly, they did not find a
signature of explosion that is distinct from steady-state convection
and/or SASI.  More recently, \citet{kotake09} used a similar
simulation approach in 3D simulations and employed a ray-tracing
technique to estimate the asymmetric neutrino emission, obtaining a
neutrino GW signal comparable in magnitude to the matter signal.
Though one might expect higher neutrino luminosities to result in
stronger convection/SASI motion, they found that the strength of the GW
emission does not correlate with neutrino luminosity.  Rather, they
concluded that the stochastic motions of the SASI prevents such a
correlation between GW strength and neutrino luminosity.

\citet{marek09a} use different approximations to investigate the GW
characteristics in 2D simulations and obtain different results.  They
use a compressible liquid-drop model EOS as well as
relativistic-mean-field and a Brueckner-Hartree-Fock EOSs, 1D ray-by-ray
neutrino transport, initial conditions derived from stellar evolution
calculations, a pseudo-GR gravitational potential that mimics the
effects of GR in spherical symmetry, and an inner boundary 
at $\sim$2 km.  With these more realistic assumptions, they obtain
matter GW amplitudes that are only half of the neutrino GW signal.
Just after bounce, they report characteristic frequencies of
$\sim$100 Hz, similar to \citet{kotake07}.  However, \citet{marek09a}
associate these frequencies with the short-lived prompt convection
rather than general SASI motions.  When non-linear SASI begins in
earnest, augmenting the nonradial motions of neutrino-driven convection, they report that the GW power peaks at $\sim$300-800 Hz and
attribute the source of GWs to vigorous SASI and convective motions
above and below the neutrinosphere.  A primary conclusion of their
work is that the peak frequency depends upon the compactness of the
PNS and, by extension, the dense-matter EOS, in that softer EOSs
result in higher frequencies.

Both groups state that GWs are generated by convective
overturns and sloshing motions of the SASI.  
In fact, \citet{kotake09} make direct reference to overturn and
sloshing timescales for the origin of GW characteristic frequencies.
\citet{marek09a} also discuss these timescales in the context of
prompt and PNS convection, but they go further to suggest that asymmetric motions in the
neutrino heating and cooling layers stirred by vigorous SASI funnels and
convective overturn are significant sources of GW emission as well.
However, neither
quantitatively define the relevant timescales, nor give
support for these associations.  In fact, most timescale definitions in the
literature associated with postshock convection and the SASI give
frequencies ($\sim$30-100 Hz) that are quite low
\citep{burrows:07b,scheck:08,marek09b} compared to the frequencies
($\gtrsim$300) associated with peak GW power.  This discrepancy
suggests that another timescale is relevant in determining the
characteristic GW frequency.  So, here we ask several
obvious questions.  What determines the characteristic frequencies and
amplitudes, how do these change with progenitor mass and neutrino
luminosity, and what is the GW signature of explosion?

Theoretical GW signals that reproduce observations with fidelity will
require 3D radiation-hydrodynamic simulations.  However, such
simulations are computationally expensive and prohibit systematic
studies of the important physics.  In the spirit of \citet{murphy08b},
we simulate simplified models that nevertheless retain the important
physics, and allow one to adjust key parameters.  Our approach is a
compromise between \citet{kotake07} and \citet{marek09a}.  Like
\citet{marek09a}, we use recent core-collapse progenitor models
\citep{woosley07} as initial conditions and a finite-temperature
relativistic-mean-field EOS.  Rather than studying the effects of the
EOS, we perform simulations with four progenitor models: 12, 15, 20,
and 40 \msun.  To avoid waiting six months (or more!) for a single
simulation, we compromise on the neutrino transport and use local
heating and cooling approximations similar to \citet{kotake07}.
Finally, we use spherical Newtonian gravity.  Though these assumptions prevent accurate models, simulations can be
calculated in a timely manner that include important qualitative
features of core collapse.  Because we do not use neutrino transport,
but a simple local heating and cooling algorithm, we calculate the GW
signal due to asymmetric mass motions and not asymmetric neutrino
emission.  Since the neutrino and matter GW signals are emitted at
low and high frequencies, respectively, and GW detectors are sensitive
to the high frequencies, calculating the matter GW signal is
sufficient for predicting GW observations in the near future.

In the following sections, we present a systematic parametrization of
progenitor mass and neutrino luminosity in the neutrino mechanism
context.
In \S \ref{section:numerics}, we present the basic equations,
assumptions, and numerical techniques.  
In \S \ref{section:results}, we present and discuss the results of this
paper.  For GW extraction, we consider the quadrupole formula of the
slow-motion, weak-field approximation (\S \ref{section:gwextract}).  In
\S \ref{section:signatures}, we describe the generic features of GW emission,
which include prompt convection, postshock convection, SASI, and
asymmetric explosions.  In \S \ref{section:gwspectra}, we show energy
spectra and spectrograms, describing the characteristic
frequencies and their dependence on progenitor mass and time of
explosion.  In \S \ref{section:gwsource}, we consider the
characteristic frequencies and amplitudes of the downdrafts striking the
PNS ``surface,'' where they decelerate due to buoyancy force and show
that these match the characteristic frequencies and amplitudes of the GWs.
Finally, in \S \ref{section:summary}, we summarize our results and
suggest a program for further investigation.

\section{Numerical Techniques}
\label{section:numerics}

\subsection{Hydrodynamics}
\label{section:hydro}

For the 2D simulations presented in this paper, we use BETHE-hydro
\citep{murphy08a,murphy08b}, an Arbitrary Lagrangian-Eulerian (ALE)
hydrodynamics code.  The basic equations of hydrodynamics are the
conservation of mass, momentum, and energy: 
\begin{equation}
\label{eq:mass_lag}
\frac{d \rho}{d t} = - \rho \nabla \cdot \bf{v} \, ,
\end{equation}
\begin{equation}
\label{eq:mom_lag}
\rho \frac{d \bf{v}}{d t} = - \rho \nabla \Phi - \nabla P
\, ,
\end{equation}
and
\begin{equation}
\label{eq:ene_lag}
\rho \frac{d \varepsilon}{d t} = - P \nabla \cdot \bf{v} 
+ \rho(\mathcal{H} - \mathcal{C}) \, .
\end{equation}
$\rho$ is the mass density, $\bf{v}$ is the fluid velocity, $\Phi$ is
the gravitational potential, $P$ is the isotropic pressure,
$\varepsilon$ is the specific internal energy, and $d/dt = \partial
/ \partial t + \bf{v} \cdot \nabla$ is the Lagrangian time
derivative.
In this work, the neutrino heating, $\mathcal{H}$, and cooling, $\mathcal{C}$, 
terms
in eq. (\ref{eq:ene_lag}) are assumed to be 
\begin{equation}
\label{eq:heating}
\mathcal{H} = 1.544 \times 10^{20} L_{\nu_e} \left ( \frac{100 {\rm
      km}}{r} \right )^2 \left ( \frac{T_{\nu_e}}{4 {\rm\, MeV}} \right
)^2 \left [ \frac{\rm erg}{{\rm g} \, {\rm s}} \right ] \, ,
\end{equation}
in which we assume the free-streaming limit,\footnote{The neutrino distribution is entirely forward peaked and the flux factor is one.} and
\begin{equation}
\label{eq:cooling}
\mathcal{C} = 1.399 \times 10^{20} \left ( \frac{T}{2 {\rm\, MeV}}
\right )^6
\left [ \frac{\rm erg}{{\rm g} \, {\rm s}} \right ] \, .
\end{equation}
Note that these approximations for heating and cooling by
neutrinos \citep{bethewilson:85,janka:01} depend upon local quantities and
predefined parameters.  They are $\rho$, temperature ($T$),
the distance from the center ($r$), the electron-neutrino temperature ($T_{\nu_
e}$), and the electron-neutrino luminosity ($L_{\nu_e}$), which
we give in units of $10^{52}$ erg s$^{-1}$.  By using
eqs. (\ref{eq:heating}) and (\ref{eq:cooling}), we gain considerable
time savings by approximating the effects of detailed neutrino transport.  For
all simulations, we set $T_{\nu_e} = 4$ MeV.  In eq.~(\ref{eq:heating}), it has
been assumed that $L_{\nu_e} = L_{\bar{\nu}_e}$ and that the mass
fractions of protons and neutrons sum to one.  Therefore, the sum of
the electron- and anti-electron-neutrino luminosities is $L_{\nu_e
  \bar{\nu}_e} = 2 L_{\nu_e}$.
Closure for eqs.~(\ref{eq:mass_lag}-\ref{eq:ene_lag}) is obtained
with the finite-temperature nuclear EOS of \cite{shen:98a} 
to which we have added the effects of photons, electrons, and positrons.
As such, the EOS has the following dependencies:
\begin{equation}
\label{eq:eos}
P = P(\rho,\varepsilon,Y_e) \, ,
\end{equation}
where $Y_e$ is the electron fraction.
Given this $Y_e$ dependence, we also solve the equation:
\begin{equation}
\label{eq:advec}
\frac{d Y_e}{dt} = \Gamma_e \, ,
\end{equation}
where $\Gamma_e$ is the net rate of $Y_e$ change.

The heating and cooling functions, eqs. (\ref{eq:heating}) \&
(\ref{eq:cooling}), are valid for optically thin regions only.  To
mimic the reduction in heating and cooling at depth where the PNS is
opaque to neutrinos, we multiply by $e^{-\tau_{\rm eff}}$. In 1D
simulations, the effective optical depth for electron- and
anti-electron-type neutrinos is calculated by
\begin{equation}
\label{eq:tau}
\tau_{\rm eff} = \int^{\infty}_r \kappa_{\rm eff}(r) dr \, ,
\end{equation}
where the effective opacity, $\kappa_{\rm eff}$ is given by eq. (15) or (16) of
\citet{janka:01}, which we reproduce without derivation:
\begin{equation}
  \label{eq:kappa}
  \kappa_{\rm eff} (r) \approx 1.5 \times 10^{-7} X_{n,p}
  \left (\frac{\rho}{10^{10} \, {\rm g} \, {\rm cm}^{-3}} \right )
  \left ( \frac{T_{\nu_e}}{4 {\rm\, MeV}} \right )^2 {\rm cm}^{-1}\, ,
\end{equation}
where $X_{n,p}$, the composition weighting, is $\sim \frac{1}{2}(Y_n + Y_p)$ and $Y_n$ and $Y_p$ are the number fractions of free neutrons and protons.
Because the radial density profile after bounce is monotonic and, to good
approximation, a series of power-laws, $\tau_{\rm eff}$ is roughly a
function of local density only.  We use our 1D simulations to calibrate
this $\tau$-$\rho$ relationship and parametrize $\tau$ as function of
$\rho$ for the 2D simulations.

The heating and cooling terms are set to zero during collapse and are
turned on after bounce.  We have experimented with several methods to
do this, but find that they give similar results provided that
these terms ``turn on'' within a few milliseconds of bounce.  For
convenience, we include heating and cooling once the maximum value of
$\tau$ is greater than 100.

Using BETHE-hydro
\citep{murphy08a}, we solve eqs.~(\ref{eq:mass_lag}-\ref{eq:ene_lag}) in two
dimensions by the Arbitrary Lagrangian-Eulerian (ALE) method. To
advance the discrete equations of hydrodynamics by one time step, ALE
methods generally use two operations, a Lagrangian hydrodynamic step
followed by a remap.  The structure of BETHE-hydro's
hydrodynamic solver is designed for arbitrary-unstructured grids, and the
remapping component offers control of the time evolution of the grid.
Taken together, these features enable the use of
time-dependent arbitrary grids to avoid some unwanted features of
traditional grids.

For the calculations presented in this paper, we use this flexibility
to avoid the coordinate singularity of spherical grids in two
dimensions.  While spherical grids are generally useful for
core-collapse simulations, the convergence of grid lines near the
center places extreme constraints on the time step via the
Courant-Friedrichs-Levy condition.  A common remedy is to simulate the
inner $\sim$10 km in 1D or to use an inner boundary condition.
Another approach, which has been used in VULCAN/2D simulations
\citep{livne:93,livne:04,burrows:07a}, is to avoid the singularity with a grid
that is pseudo-Cartesian near the center and smoothly transitions to a
spherical grid at larger radii.

In this paper, we use a similar grid, the butterfly mesh
\citep{murphy08a,murphy08b} interior to 34 km and a spherical grid
exterior to this radius.  See
\citet{murphy08a} for an example.  For all 180$^{\circ}$ simulations,
the innermost pseudo-rectangular region is 50 by 100 zones, and the
region that transitions from Cartesian to spherical geometry has 50
radial zones and 200 angular zones.  The outermost spherical region
has 200 angular zones and 500 radial zones which are logarithmically
spaced between 34 km and 8000 km in most cases.  Due to EOS
constraints, the outer boundary goes out to 6000 km for the 12-\msun\
model.  The butterfly portion has an effective radial resolution of
0.34 km with the shortest cell edge being 0.28 km.  In 2D simulations,
we do not remap every time step.  Since the time step is limited by
the large sound speeds in the PNS core, we can afford to perform
several Lagrangian hydrodynamic solves before remapping back to the
original grid.  By remapping rarely, we gain considerable savings in
computational time.

We make several approximations to expedite the calculations.  For one,
gravity is calculated via $\vec{g} = -G M_{\rm int}/r^2$, where
$M_{\rm int}$ is the mass interior to the radius $r$. Furthermore,
since we do not solve the neutrino transport equations, we do not
obtain electron/positron capture and emission rates,
which are necessary for self-consistent $Y_e$ profiles and evolution.
Instead, we use a $Y_e$ vs.\ $\rho$ parametrization that
is quantitatively accurate during collapse and gives the correct
qualitative trends after bounce.
\citet{liebendorfer05a} observed that 1D simulations including
neutrino transport produce $Y_e$ values during collapse which are
essentially a function of density alone.  This allows for a
parametrization of $Y_e$ as a function of $\rho$.  To change $Y_e$, we
use results of 1D SESAME \citep{burrows:00,thompson:03} simulations to
define the function $Y_e(\rho)$, and we employ the prescription of
\citet{liebendorfer05a} to calculate local values of $\Gamma_e$.
Though this prescription is most appropriate during collapse, we also
apply it after bounce to continue deleptonizing postshock material at
later times.  While this is not entirely consistent with our heating
and cooling prescription, the most important effects of
deleptonization are included without the need for expensive, detailed
neutrino transport.

The approximations for neutrino-matter interactions, electron capture
rates, and gravity give simulations that reproduce the primary
features of the core-collapse problem.  These include a PNS, a stalled
shock at $\sim$200 km, a gain radius at $\sim$100 km (that divides the
gain region above with net heating from the cooling region below), prompt
convection, PNS convection from $\sim$20 to $\sim$45 km, postshock
convection, and the SASI.  While the $Y_e$-$\rho$ parametrization is
designed to give the proper $Y_e$ profile during collapse, we do not
reproduce the $Y_e$ trough at later times that extends from $\sim$40
km to the shock (e.g., \citealt{dessart:06a}).  \citet{dessart:06a}
conclude that the negative lepton gradient from $\sim$15 to $\sim$30
km in simulations employing consistent neutrino transport drives inner
the
PNS convection.  In our simulations, PNS convection is driven by a
negative entropy gradient.  Although 1D simulations that solve
neutrino transport more consistently show a similar negative entropy
gradient, the depth and width are smaller in those simulations
\citep[compare Figs. 13 and 6 of][]{murphy08b,liebendoerfer:05}.  Furthermore, our PNS convection velocities
are roughly a factor of two larger than those obtained by
\citet{dessart:06a}.  In summary, while we obtain a PNS convective
region, our PNS convection is predominantly driven by an entropy
gradient rather than by a negative lepton gradient and is more
extended and vigorous compared to more realistic calculations.
However, we show in later sections that though the GW signal due to PNS
convection is important, it is not an overwhelmingly dominant feature of
GW emission from CCSNe.  Therefore, the differences in PNS convection should
not change the qualitative conclusions of this paper.

\subsection{Progenitor Models}
\label{section:progenitor}

As initial conditions, we use the 12-, 15-, 20-, and 40-\msun\ core-collapse progenitor models of \citet{woosley07}, where
these masses correspond to the zero-age-main-sequence (ZAMS) masses.
In Fig. \ref{hd_lumspanel}, we plot the mass accretion rate vs. time
after bounce (solid black lines) for each of these models.  In
general, because the more massive progenitors have extended core
structures, the accretion rates for them are higher at a given time.
Since higher accretion rates require larger neutrino luminosities for
successful explosions \citep{murphy08b}, this translates into higher
critical neutrino luminosities for the more massive progenitors.  In fact, if we assume that only the neutrino mechanism leads
to explosion, the
required neutrino luminosity for successful explosions of the 40-\msun\ progenitor is so high, $\gtrsim 10^{53}$ erg s$^{-1}$, we doubt
that it will explode at all and suspect that the non-explosive
simulation using $L_{\nu_e} = 6 \times 10^{52}$ erg s$^{-1}$
represents the likely outcome of core collapse for the 40-\msun\ progenitor.  In
general, these trends with progenitor mass will translate to trends in
the GW signal which we discuss in forthcoming sections.

Although core structure follows a general trend with the progenitor's
ZAMS mass, the correlation is known to be somewhat chaotic \citep{wwh:02}. 
For example, a substantial drop in the 20-\msun\ model's accretion rate corresponding to the interface
between the Si and O burning shells temporarily reduces the accretion
below even the 12- and 15-\msun\ progenitors at $\sim$300 ms past
bounce.  This has consequences for the development of the SASI,
explosion, and the GW signal.

\section{Results}
\label{section:results}

\subsection{Gravitational Wave Extraction and Analysis}
\label{section:gwextract}

We extract the gravitational wave signal from our hydrodynamic
simulations via the slow-motion, weak-field formalism (e.g.,
\citealt{mtw}) and consider the dominant mass-quadrupole only.  In
this approximation, the gravitational wave field is directly
proportional to the transverse-traceless (TT) part of the second time
derivative of the reduced mass-quadrupole tensor
$\Ibar_{jk}$. Specifically,
\begin{equation}
h^{TT}_{jk} (t,{\bf x}) = \frac{2}{c^4}\frac{G}{D}
\bigg[\frac{d^2}{dt^2}\, \Ibar_{jk} (t - D/c)\bigg]^{TT}\,\,,
\label{eq:sqf2}
\end{equation}
and
\begin{equation}
\Ibar_{jk} = \int \rho 
\bigg(x^j x^k - \frac{1}{3}\delta^{jk} x_i x^i\bigg) d^3 x\,.
\label{eq:Ibar}
\end{equation}
In the above, $D$ is the observer distance to the source of emission.
All other variables and constants have their usual meaning and the
Einstein sum convention is assumed. For numerical convenience,
we employ the so-called first-moment-of-momentum-divergence (FMD) recast
of eq.~(\ref{eq:sqf2}) proposed by \cite{finnevans:90}, which employs
the continuity equation to remove one time derivative.
In axisymmetry, $\Ibar_{ij}$ reduces to one independent component
($\Ibar_{zz}$) and, adapted to spherical coordinates, the FMD
formula for its first time derivative reads
\begin{eqnarray}
\label{eq:dibardt}
\frac{d}{dt}\, \Ibar_{zz} &=& \frac{8\pi}{3} \int_{-1}^{1} d\cos\theta \int_{r_1}^{r_2} dr\, r^3 \rho \, \, \cdot\nonumber \\
&&\bigg[ P_2(\cos\theta)\, v_r + \frac{1}{2} \frac{\partial}{\partial \theta} P_2(\cos\theta)\, v_\theta \bigg]\,\,,
\end{eqnarray}
where $P_2(\cos\theta)$ is the second Legendre polynomial in
$\cos\theta$ and $v_r$ and $v_\theta$ are the fluid velocities in the
radial and lateral direction, respectively. The
axisymmetric\footnote{Note that in axisymmetry $h_\times = 0$
  everywhere and $h_+$ is non-zero only away from the axis of symmetry
  of the system.}  gravitational wave strain $h_+ = h_{\theta\theta}$
(dropping the TT superscript) is then given by
\begin{equation}
h_+ = \frac{3}{2}\frac{G}{Dc^4} \sin^2 \alpha \frac{d^2}{dt^2}\,\, \Ibar_{zz}\,\,,
\label{eq:fmd}
\end{equation}
where $\alpha$ is the angle between the symmetry axis and the line of
sight of the observer and the second time derivative is obtained via
straightforward differentiation\footnote{Note that due to a  misprint
in Eq.~(38) of \citealt{finnevans:90} their expression is a factor of
2 smaller than our expression in Eq.~\ref{eq:fmd}.}.
For our simulations, this approach
yields good results and we do not find it necessary to employ the
modification of eq.~(\ref{eq:sqf2}) proposed by \cite{blanchet:90} that
removes both time derivatives, but introduces a sensitive dependence
on derivatives of the Newtonian gravitational potential, which we treat
only in spherical fashion.


In our analysis of the GW signature, we not only consider the
dimensionless GW strain $h_+$, but also compute the total emitted
energy in GWs, given by
\begin{equation}
E_\mathrm{GW} = \frac{3}{10} \frac{G}{c^5} \int_0^t 
\bigg(\frac{d^3}{dt^3}\, \Ibar_{zz}\bigg)^2\,dt\,,
\label{eq:egw}
\end{equation}
and the GW spectral energy density via 
\begin{equation}
\frac{d E_\mathrm{GW}}{df} = \frac{3}{5}\frac{G}{c^5} (2\pi f)^2 |\tilde{A}|^2\,,
\label{eq:spect}
\end{equation}
where $\tilde{A}$ denotes the Fourier transform of $A \equiv
\frac{d^2}{dt^2}\, \Ibar_{zz}$ computed via
\begin{equation}
\tilde{A}(f) = \int_{-\infty}^{\infty} A(t)\, e^{-2\pi i f t} dt\,\,.
\end{equation}
At this point, it is useful to define for future reference the
dimensionless characteristic gravitational wave strain
\citep{flanhughes:98}, in terms of the GW spectral energy density,
\begin{equation}
\label{eq:hchar}
h_\mathrm{char} = \sqrt{\frac{2}{\pi^2} \frac{G}{c^3} 
\frac{1}{D^2} \frac{dE_\mathrm{GW}}{df}}\,\,.
\end{equation}

For signals with relatively stable frequencies and amplitudes, Fourier
transforms and their energy spectra are adequate frequency analysis
tools.  However, for signals with time-varying amplitudes and
frequencies, a short-time Fourier transform (STFT) is more
appropriate.  The STFT of $A(t)$ is
\begin{equation}
\label{eq:stft}
\tilde{S}(f,\tau) = \int_{-\infty}^{\infty} A(t)\, H(t - \tau) \,
e^{-2\pi i f t} dt\,\, ,
\end{equation}
where $\tau$ is the time offset of the window function, $H(t-\tau)$.
We use the Hann window function:
\begin{equation}
\label{eq:hann}
H(t-\tau) = 
\left \{
\begin{array}{lcl}
\frac{1}{2} \left ( 1 + \cos \left ( \frac{\pi (t -
  \tau)}{\delta t}\right ) \right ) & \mbox{for} & |t-\tau| \le
  \frac{\delta t}{2}\\
0 & \mbox{for} & |t-\tau| > \frac{\delta t}{2}\\
\end{array}
\right.
 \, ,
\end{equation}
where $\delta t$ is the width of the window function.
The analog of the energy spectrum of the Fourier
transform is the spectrogram, $|\tilde{S}(f,\tau)|^2$.  Using the
spectrogram, we define an analog to the energy emission per frequency
interval (eq. \ref{eq:spect}):
\begin{equation}
\label{eq:spectrogram}
\frac{d E^*_\mathrm{GW}}{df}(f,\tau) = \frac{3}{5}\frac{G}{c^5} (2\pi f)^2
|\tilde{S}(f,\tau)|^2 \, .
\end{equation}

We emphasize that the GW strains reported in this paper are based upon
matter motions alone and do not include the low frequency signal that
results from asymmetric neutrino emission \citep{burrows96,mueller97}.
Accurate calculations of asymmetric neutrino emission require multi-D,
multi-angle neutrino transport to capture the true asymmetry of the
neutrino radiation field (see, e.g., \citealt{ott:08}). Our choice to
parametrize the effects of neutrino transport by local heating and
cooling algorithms is based upon assumptions of transparency, which
ignore diffusive effects and would exaggerate the asymmetries and
resulting GWs.  For example, \citet{kotake07} estimated the neutrino
GW signal using a similar heating and cooling parametrization and
obtained GW strain amplitudes that are $\sim$100 times the matter GW
signal.  However, with an improved ray-tracing-based method, the same
authors find much smaller amplitudes that are larger than those due to
matter motions by only a factor of a few \citep{kotake09}. This is in
agreement with the GW estimates of \citet{marek09a} who used 1D
ray-by-ray neutrino transport and coupled neighboring rays in 2D
hydrodynamic simulations.

Studying the matter GW signal alone is worthwhile.  Although the
neutrino GW strain amplitudes can be as large or even larger than the
contribution by matter
\citep{burrows96,mueller97,mueller:04,marek09a}, the typical
frequencies, $f$, of the neutrino GW signal ($\sim$10 Hz or less) are
typically much lower than the frequencies of the matter signal
($\gtrsim$100 Hz).  Consequently, the GW power emitted, which is
proportional to $f^2$, can be much higher for the matter GW signal.
Furthermore, although future GW detectors (e.g. Advanced LIGO) will have
improved sensitivity at low frequencies, current detectors have
response curves that are not sensitive to the lower frequencies of the
neutrino GW signal.

\subsection{Signatures in the GW Strain}
\label{section:signatures}

\begin{figure}
\epsscale{0.7}
\plotone{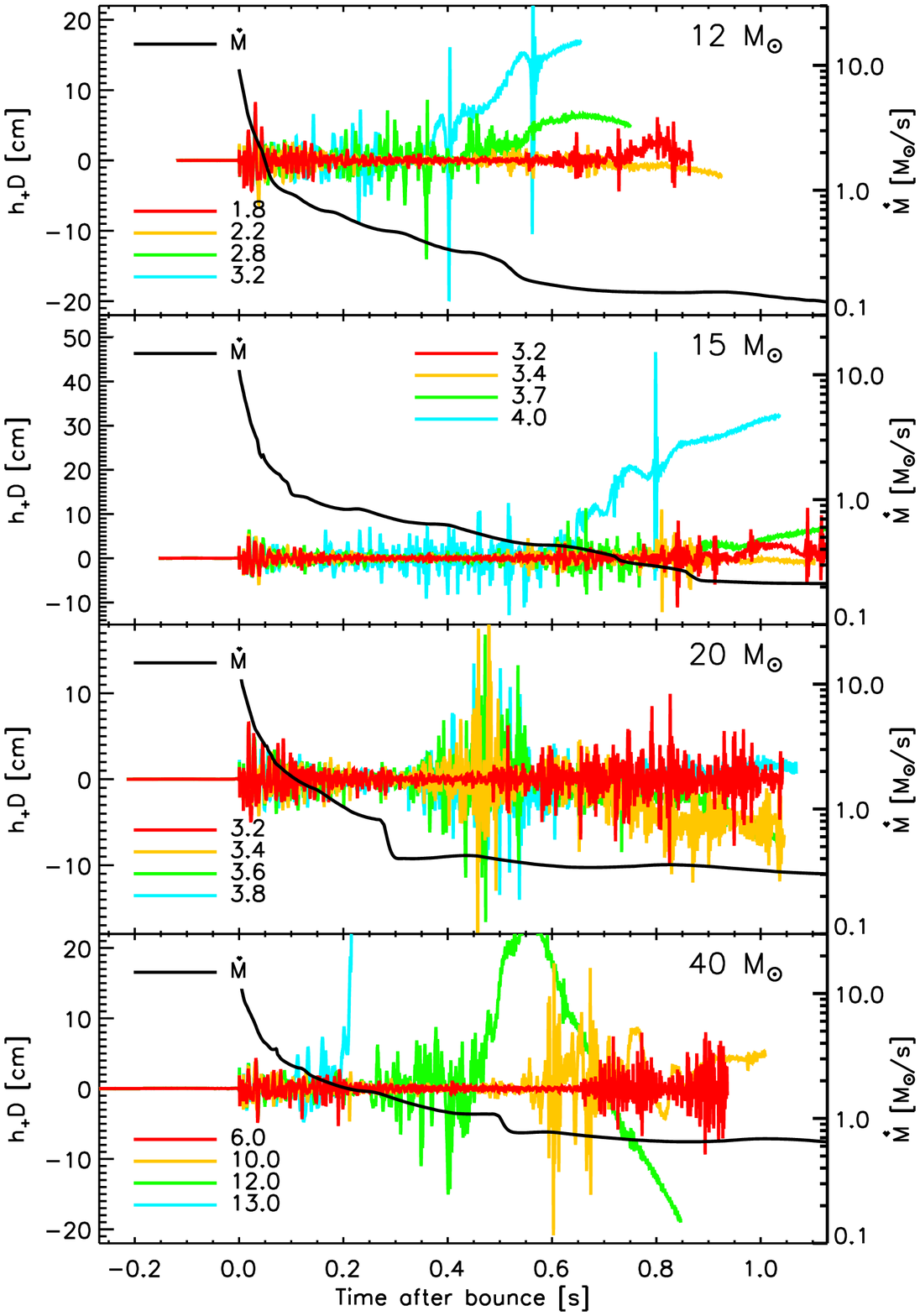}
\caption{The dimensionless GW strain $h_+$ times the distance, $D$, vs. time after bounce for the suite of simulations
  presented in this paper.  Each panel represents a single progenitor
  model and the curves are labeled by $L_{\nu_e}$ in units of
  $10^{52}$ erg s$^{-1}$.  The accretion rates (in \msun\ s$^{-1}$) at the
  stalled shock of non-exploding models are shown for comparison.  All
  models show GW features associated with prompt convection, postshock
  convection, SASI, and explosion with the exception that the run
  using $M = 40$ \msun\ and $L_{\nu_e} = 6.0 \times 10^{52}$ erg s$^{-1}$ does not explode.  See
  Fig. \ref{hdlabeled} for a sample GW strain evolution with these features
  labeled.  In general, lower luminosity runs develop the strong SASI
  motions and explosions at later times, but all runs saturate at
  roughly similar amplitudes ($h_+D \sim 10$ cm).  In the simulations
  using the 20-\msun\ progenitor, the accretion of the interface
  between the progenitor's Si and O burning shells through the shock
  causes a sudden drop in accretion rate.  This initiates strong
  convective/SASI motions in the highest luminosity runs (3.4, 3.6,
  and 3.8) and suggests a means to use GW signals to probe the
  location of this interface.  However, this feature is unique to the
  runs using the 20-\msun\ progenitor. \label{hd_lumspanel}}
\epsscale{1.0}
\end{figure}

\begin{figure}
\plotone{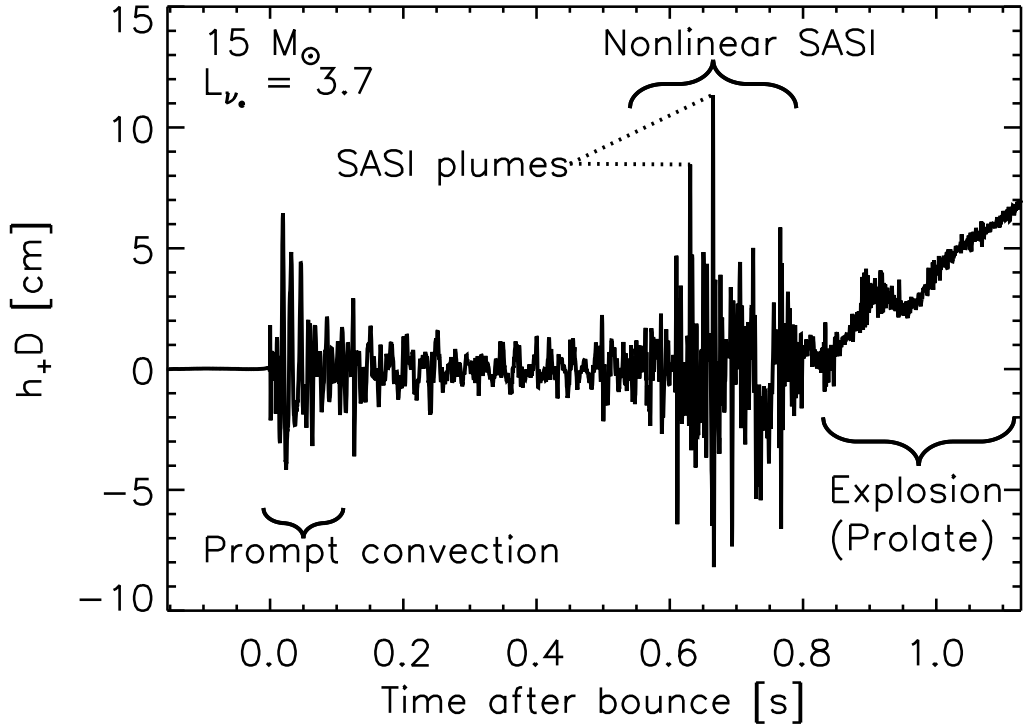}
\caption{A sample of GW strain ($h_+$) times the distance, $D$, vs. time after bounce.  This signal was
  extracted from a simulation using a 15-\msun\ progenitor model
  \citep{woosley07} and an electron-type neutrino luminosity of
  $L_{\nu_e} = 3.7 \times 10^{52}$ erg s$^{-1}$.  Prompt convection,
  which results from a negative entropy gradient left by the stalling
  shock, is the first distinctive feature in the GW signal from 0 to
  $\sim$50 ms after bounce.  From $\sim$50 ms to $\sim$550 ms past
  bounce, the signal is dominated by PNS and postshock convection.
  Afterward and until the onset of explosion ($\sim$800 ms), strong
  nonlinear SASI motions dominate the signal.  The most distinctive
  features are spikes that correlate with dense and narrow down-flowing
  plumes striking the ``PNS'' surface ($\sim$50 km).  Around $\sim$800
  ms the model starts to explode.  In this simulation, the
  GW signal during explosion is marked by a significant decrease in
  nonlinear SASI characteristics.  The aspherical (predominantly
  prolate) explosion manifests in a monotonic rise in $h_+D$ that is
  similar to the ``memory'' signature of asymmetric neutrino emission.
  \label{hdlabeled}}
\end{figure}

In Fig.~\ref{hd_lumspanel}, we plot the GW strain (eq.~\ref{eq:fmd})
times the distance to a 10 kpc source, $h_+D$, vs. time after bounce
for all simulations.  Though there is some diversity in amplitude and
timescale among these GW strains, there are several recurring
features that exhibit systematic trends with mass and neutrino
luminosity.  We illustrate these features in Fig.~\ref{hdlabeled} with
the GW strain of the simulation using the 15-\msun\ progenitor and
$L_{\nu_e} = 3.7 \times 10^{52}$ erg s$^{-1}$.  Before bounce, spherical collapse results in zero
GW strain.  Just after bounce the prompt shock loses energy and
stalls, leaving a negative entropy gradient that is unstable to
convection.  Because the speeds of this prompt convection are larger
than those of steady-state postshock or PNS convection afterward, the
GW strain amplitude rises to $h_+D \sim 5$ cm during prompt convection
and settles down to $\sim$1 cm roughly 50 ms later, which is consistent
with the results of \citet{ott:09a} and \citet{marek09a}.  Later in
this section, we show that during both phases, convective motions in
postshock convection above the neutrinosphere and PNS convection below
it contribute to the GW strain.  Since nonlinear SASI oscillation amplitudes
increase around 550 ms past bounce, the GW signal strengthens from
$h_+D \sim 1$ to 10 cm and is punctuated by spikes that are coincident
in time with narrow plumes striking the PNS ``surface'' (at $\sim$50
km).  \citet{marek09a} also noted this correlation.  

\begin{figure}
\plotone{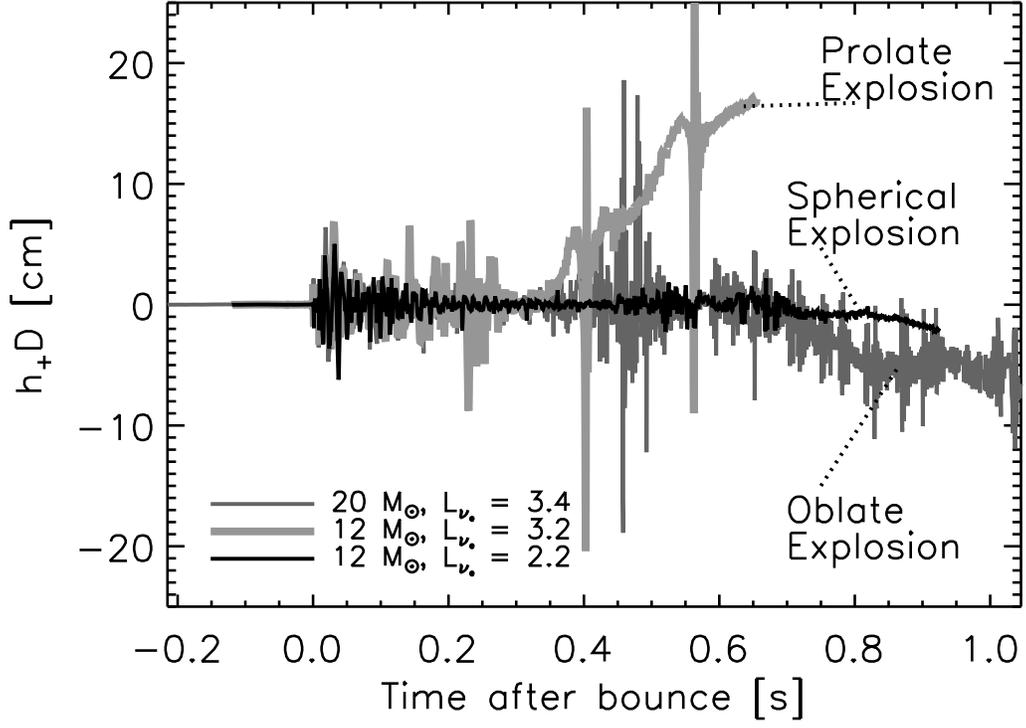}
\caption{Plots of $h_+D$ vs.\ time after bounce for three models showing the
  effect of global asymmetries in explosion.  The general
  shapes of the explosions are prolate ($M = 12$ \msun, $L_{\nu_e} =
  3.2 \times 10^{52}$ erg s$^{-1}$, and thick light-gray line), spherical ($M = 12$ \msun, $L_{\nu_e} =
  2.2 \times 10^{52}$ erg s$^{-1}$, and thin black line), and oblate ($M = 20$ \msun, $L_{\nu_e} =
  3.4 \times 10^{52}$ erg s$^{-1}$, and thin gray line), and correspond to positive, zero, and
  negative ``memory,'' respectively.  See
  Fig.~\ref{prolateoblatestills} for color maps of the entropy
  distribution during explosion for the same models.  \label{prolateoblatehd}}
\end{figure}

The final feature after $\sim$800 ms is associated with explosion.
The signatures of explosion are twofold.  First, during explosion,
postshock convection and the SASI subside in strength and the higher
frequency ($\sim$300-400 Hz) oscillations in $h_+D$ diminish.  Second,
global asymmetries in mass ejection result in long-term and large
deviations of the GW strain.  In Fig. \ref{hdlabeled}, a monotonic
rise of $h_+D$ to nonzero, specifically positive, values corresponds
to a prolate explosion in this simulation. This is similar to the
``memory'' in the GW signal of asymmetric neutrino emission
\citep{burrows96,mueller97}.  When the explosion is spherical, the strain drops
to zero and remains there, and when it is oblate, the strain maintains
negative values.  Examples of $h_+D$ curves showing prolate, oblate,
and spherical explosions are shown in Fig. \ref{prolateoblatehd}.  The
simulation using a 20-\msun\ progenitor and $L_{\nu_e} = 3.4 \times
10^{52}$ erg s$^{-1}$ (gray
line) exploded with an oblate structure, the simulation with $M = 12$
\msun\ and $L_{\nu_e} = 3.2 \times 10^{52}$ erg s$^{-1}$ (light gray line) has a prolate
explosion, and the simulation with $M = 12$ \msun\ and $L_{\nu_e} =
2.2 \times 10^{52}$ erg s$^{-1}$ explodes almost spherically.

\begin{figure}
\plotone{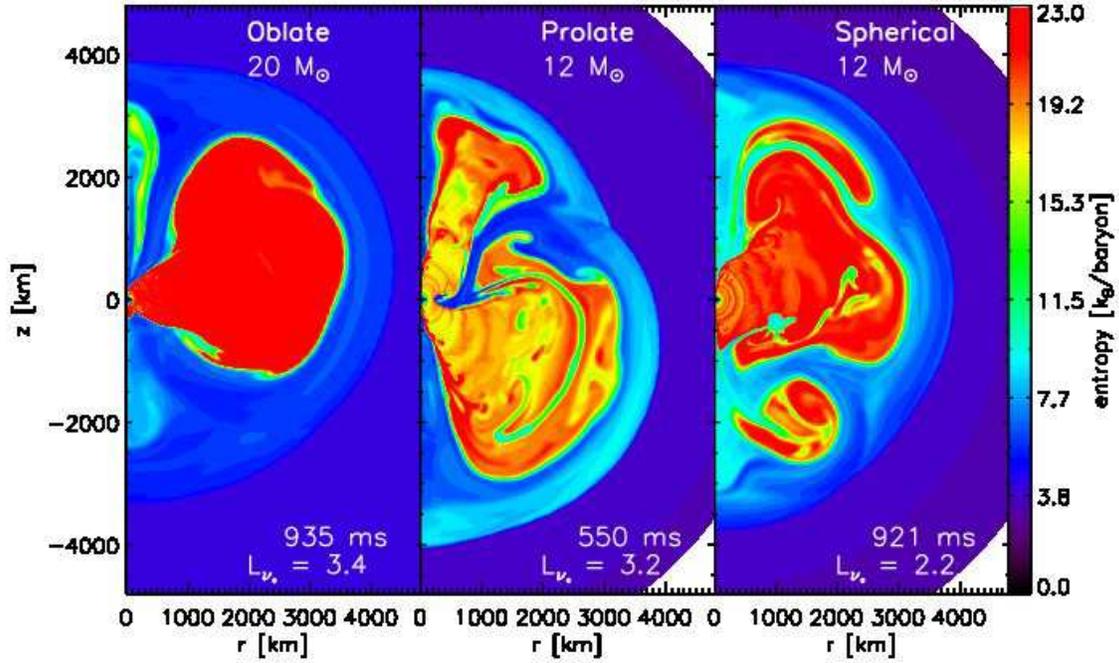}
\caption{Snapshots of the entropy ($k_B$/baryon) distribution during explosion for
  the three models highlighted in Fig. \ref{prolateoblatehd}.  Lighter shades
  (warmer colors in the online version) represent higher entropies, while darker shades (cooler colors in the online version) shades represent
  lower entropies.  The global asymmetry of the matter interior to the
  shocks is oblate ($M = 20$ \msun\ and $L_{\nu_e} = 3.4 \times
  10^{52}$ erg s$^{-1}$), prolate ($M
  = 12$ \msun\ and $L_{\nu_e} = 3.2 \times 10^{52}$ erg s$^{-1}$), and spherical ($M = 12$ \msun,
  $L_{\nu_e} = 2.2 \times 10^{52}$ erg s$^{-1}$).  Even though the
  GW signal may indicate that the explosion is in general
  ``spherical,'' ``oblate,'' or ``spherical,'' the
  entropy (hence temperature and composition) distributions may show
  higher order structure. \label{prolateoblatestills}}
\end{figure}

Figure \ref{prolateoblatestills} shows snapshots of the entropy (in
units of $k_B$/baryon) distribution during explosion for the three
models highlighted in Fig.~\ref{prolateoblatehd}.  Lighter shades (warmer colors in the online version)
represent higher entropies, while darker shades (cooler colors in the online version) represent lower
entropies.  The general shapes of the matter interior to the shocks is
oblate ($M = 20$ \msun\ and $L_{\nu_e} = 3.4 \times 10^{52}$ erg s$^{-1}$), prolate ($M = 12$
\msun\ and $L_{\nu_e} = 3.2 \times 10^{52}$ erg s$^{-1}$), and spherical ($M = 12$ \msun, $L_{\nu_e} =
2.2 \times 10^{52}$ erg s$^{-1}$; and thin black line).  We emphasize that the GW signal is
sensitive to $\ell = 2$ accelerations of matter and is somewhat blind
to differences in composition or higher order asymmetries.  Therefore,
even though the GW signal may indicate that the explosion is in
general ``spherical,'' ``oblate,'' or ``prolate,'' the entropy
(hence temperature and composition) distributions may not be.

\begin{figure}
\epsscale{1.0}
\plotone{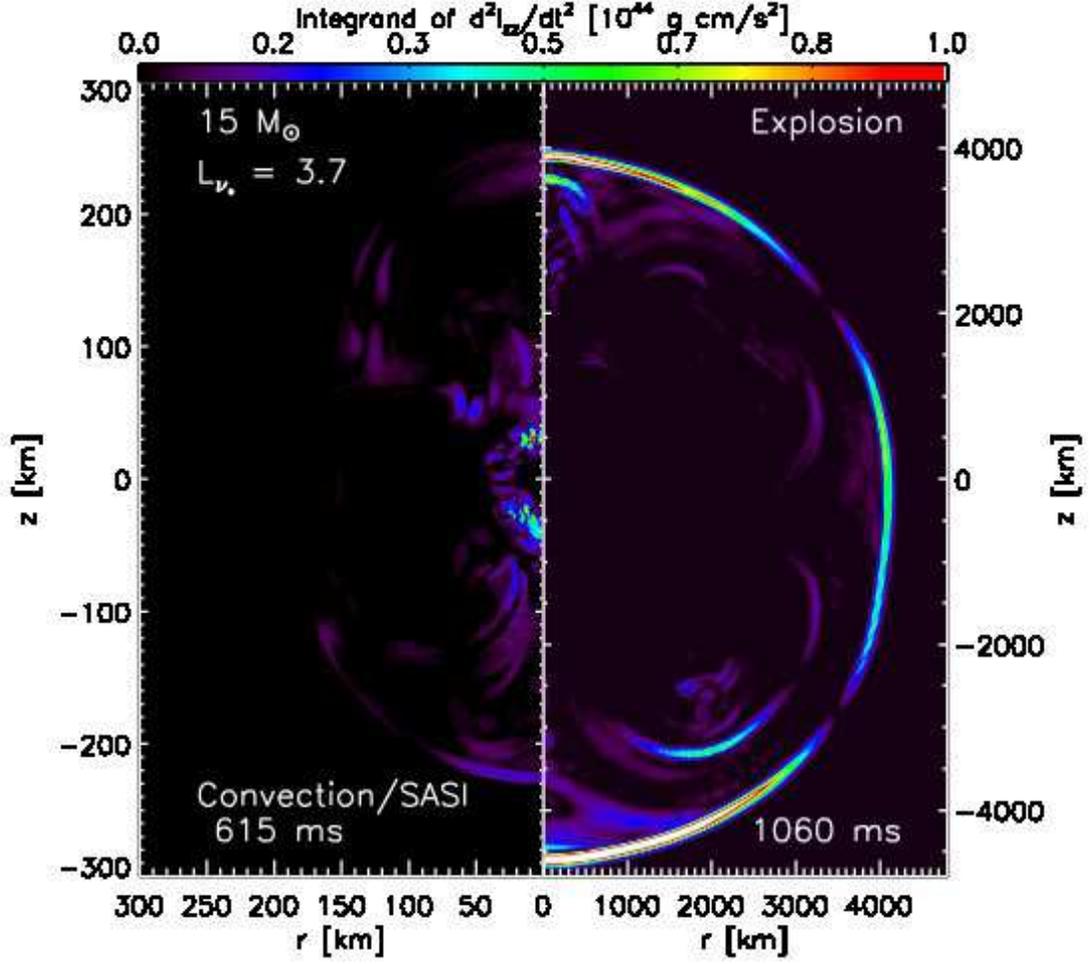}
\caption{Gray-scale (color in the online version) map of the integrand that leads to the GW
  signal (See eqs. \ref{eq:dibardt}, \ref{eq:fmd}, and
  \ref{eq:integrand}).  Lighter shades (brighter colors in the online version) correspond to higher values.
  The results are for the 15-\msun\ progenitor and $L_{\nu_e} = 3.7
  \times 10^{52}$ erg s$^{-1}$.  The left panel represents the
  stalled-shock phase during vigorous postshock convection/SASI
  motions.  During this phase, the signal originates predominantly from
  PNS convection and the deceleration of plumes below the gain radius
  ($\sim$100 km).  The right panel shows that the ``memory'' signal
  during explosion is a result of acceleration at an asymmetric shock.
  \label{hdintegrandmaps}}
\epsscale{1.0}
\end{figure}

\begin{figure}
\epsscale{0.8}
\plotone{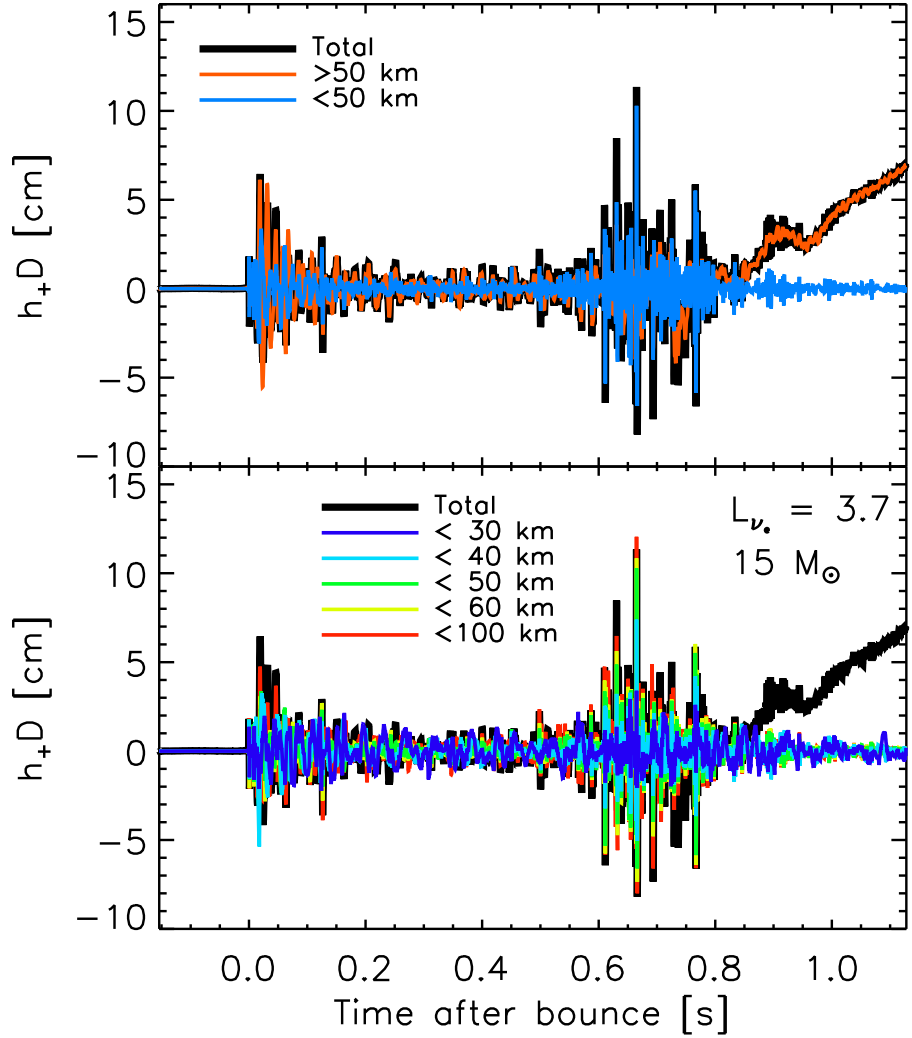}
\caption{GW waveforms, $h_+D$ vs. time, showing the contributions of
  PNS convection and the SASI.  The model shown is the same as in
  Fig.~\ref{hdlabeled}.  For reference, the entire GW signal is shown
  in both panels (solid-black line).  The signal originating from $>
  50$ km (orange, online version) and $< 50$ km (blue, online version) is shown in the top panel.
  This radius is roughly the division between nonlinear SASI motions
  and PNS convection motions.  Most, but not all, of the signal
  associated with prompt convection originates in the outer convection
  zone.  There is a non-negligible contribution from the PNS.  The
  monotonic rise in the GW strain during explosion clearly originates
  from the outer (exploding) regions.  Even though the region for
  the convection/SASI and its nonlinear motions are above 50 km, these motions
  influence the PNS convective motions below 50 km.  It is telling
  that once the model explodes, and the nonlinear SASI motions subside
  but the PNS convection does not, the GW signal from below 50 km
  diminishes as well.  The bottom panel shows the GW signal from five
  regions, each with different outer radii (30, 40, 50, 60, and 100 km).
  The strengthening of the GW signal associated with the SASI is
  apparent for all, suggesting that the influence of the SASI
  diminishes only gradually with depth.
  \label{hd_totpnssasi}}
\epsscale{1.0}
\end{figure}

In Figs. \ref{hdintegrandmaps} and \ref{hd_totpnssasi}, we localize
the source of GW emission.  Figure \ref{hdintegrandmaps} shows the
spatial distribution of
\begin{equation}
\label{eq:integrand}
\left |
\frac{d}{dt}
\left [
r^3 \rho \left ( P_2(\cos\theta)\, v_r + \frac{1}{2}
\frac{\partial}{\partial \theta} P_2(\cos\theta)\, v_\theta \right ) 
\right ]
\right |
\,,
\end{equation}
which is the time derivative of the integrand of
eq. (\ref{eq:dibardt}) and determines the GW strain
(eq. \ref{eq:fmd}).  As in Fig. \ref{hdlabeled}, these results are for
the 15-\msun\ progenitor and $L_{\nu_e} = 3.7 \times 10^{52}$ erg
s$^{-1}$.  Lighter shades (brighter colors in the online version) correspond to a larger contribution to the GW
strain integral.  The left panel is at 615 ms past bounce, and
represents the phase with a stalled shock ($\sim$250 km) and vigorous
postshock convection/SASI motions.  This is our first indication that
motions below the gain radius ($\sim$100 km), in particular
deceleration of plumes and PNS convection, are the strongest sources
of GWs before explosion.

The right panel
is at 1060 ms past bounce and shows that accelerations at an
aspherical shock ($\sim$4000~km at this time) are responsible for the
``memory'' signature at late times.  Using the appropriate shock
velocities and asymmetries and postshock densities and velocities in
eqs. (\ref{eq:dibardt}) and (\ref{eq:fmd}), we obtain the correct
order-of-magnitude for the memory signal.  However, during explosion a
high entropy wind emerges that encounters the swept-up material of the
primary shock and produces secondary shocks.  We find that these
secondary shocks and their asymmetric structure can add a non-trivial
contribution to the ``memory'' signal.  The strength of these winds is
a strong function of the neutrino luminosity
\citep{qian96,thompson01}.  Since we employ a constant neutrino
luminosity, an accurate characterization of the late-time ``memory''
(such as saturation) are beyond the scope of this paper.

Figure~\ref{hd_totpnssasi} quantifies the sources of GW radiation and
their spatial distribution as a function of radius.  The model shown is the
same as in Figs. \ref{hdlabeled} and \ref{hdintegrandmaps}.  For reference, the entire GW signal
is shown in both panels (solid-black line).  The signals originating
from $> 50$ km (orange, online version) and $< 50$ km  (blue, online version) are shown in the top panel.
Below 50 km, PNS convection dominates the motions, and above this
radius postshock convection and the SASI dominate.  As expected,
most of the signal associated with prompt convection originates in the
outer convective zone, though motions below 50 km and presumably from
PNS convection account for a fair fraction ($\sim$20\%).  Afterward,
the contributions to the GW amplitude from below and above 50 km are
comparable.  This suggests that motions associated with both PNS
convection and the postshock-convection/SASI region contribute
significantly to the GW signal.  The
``memory'' signature of the GW strain during explosion clearly
originates from the outer (exploding) regions.  Interestingly, once the model
explodes, and the nonlinear postshock convection/SASI motions subside, the GW signal from
below 50 km diminishes as well, though PNS convective motions do not.

To further understand the origin of GW emission, the bottom
panel of Fig. \ref{hd_totpnssasi} refines the spatial
origin of the GW emission into five regions.  We
plot the GW strain from five overlapping regions, and each extends
from the center to different outer radii (30, 40, 50, 60, and 100 km).
PNS convection extends from $\sim$20 to $\sim$40 km and the gain
radius is $\sim$100 km.  However, as we explain in section \S
\ref{section:gwsource}, the 
turbulent motions of postshock convection/SASI penetrate below the
gain radius to radii of $\sim$ 60 km during the most vigorous phases.
Therefore, the partial GW signals with outer radii of 30 and 40 km
contain signals only due to PNS convection; 50 km encompasses PNS
convection and gravity waves that are excited by the overlying
convection/SASI; 60 km encompasses PNS convection, gravity waves, and
the deepest penetration of the convection/SASI plumes (see \S
\ref{section:gwsource}); and 100 km encompasses all of these
contributions and the gain radius.  First, we note that extending the
outer radius from 60 to 100 km, the 
gain radius, adds very little signal during the most
vigorous postshock convection/SASI phase from 550 to 800 ms past
bounce.  Therefore, the most relevant motions are those at $\sim$60 km
and below, which include the postshock convection/SASI plume
decelerations, excited gravity waves, and PNS convection.
Furthermore, the strength of the GW signal from all
radii increases and decreases in synchrony with the vigor of
postshock convection/SASI motions, which are restricted to radii above
$\sim$50 km.  This suggests that the influence of the convection/SASI
diminishes only gradually with depth.  Even though the mechanism for
convection/SASI and their nonlinear motions are above 50 km, they
influence PNS convective motions below 50 km, which results in GW
emission whose vigor is ultimately due to the postshock
convection/SASI motions.

Hence, we conclude that features of the SASI and postshock convection,
in particular the plumes striking the PNS ``surface'', are ultimately
responsible for the strong GW signal from $\sim$550~ms until
explosion at $\sim$800~ms after bounce.  Some of the signal comes
from the accelerations at this interface.  The rest comes from gravity
waves and motions
within the PNS ($< 50$ km) that are excited by the plumes striking
this layer.  Whether these motions are excited g-waves, enhanced PNS
convection, or a combination of both is difficult to distinguish in
our simulations.  A focused investigation on this aspect is warranted.

\citet{mueller:04} conclude that the GW signal from PNS
convection is a few factors smaller in amplitude than the GWs from
postshock convection.  At first sight, this conclusion appears at odds with our results
that the signals above and below 50 km are roughly equal.  However, to
calculate the GW signal from PNS convection, \citet{mueller:04} used results
from a simulation that extended
out to only 60 km \citep{keil96}, which ignores any influence that
postshock-convective or SASI motions have on PNS convection.  This is
consistent with the GW signal we obtain after explosion, when
postshock convection and SASI go away, but PNS convection remains.
These results emphasize the importance of calculating PNS convection
in the full context of core-collapse dynamics.

We now return to Fig.~\ref{hd_lumspanel} and analyze the trends with
progenitor mass and $L_{\nu_e}$.  Each panel represents a single
progenitor model and the curves are labeled by $L_{\nu_e}$.  The
accretion rates (in \msun\ s$^{-1}$) at the stalled shock of non-exploding
models are shown for comparison.  The prominent features described in
Fig.~\ref{hdlabeled} are apparent in all models, except the run using
$M = 40$ \msun\ and $L_{\nu_e} = 6.0 \times 10^{52}$ erg s$^{-1}$ which does not explode and manifests
the larger amplitudes associated with the nonlinear SASI at much later
times.  Otherwise, lower-luminosity simulations take longer to develop
the nonlinear-SASI GW signal and explode at later times, but all runs seem
to saturate at roughly similar amplitudes.  This is consistent with
the results of \citet{murphy08b}, in which they note that lower
luminosity runs take longer to reach saturated shock oscillations, but
that the saturated amplitudes are similar for the different neutrino
luminosities.  Similarly, \citet{kotake09} conclude that the GW
amplitudes are independent of neutrino luminosity.  Finally, we note
that the significant drop in accretion rate in simulations using the
20-\msun\ progenitor, instigates strong nonlinear SASI motions and
associated GW signatures.  This might suggest using GWs as a
diagnostic for the location of shell burning or compositional interfaces.  However,
very few of the other simulations show such a convincing correlation.

The neutrino luminosities required to explode the 40-\msun\ progenitor
model are much larger than simulations obtain that consistently
calculate neutrino transport \citep{dessart:06a,ott:08,marek09b}.
Therefore, if other mechanisms fail to explode higher mass stars, we suspect that the simulation using $L_{\nu_e} = 6 \times
10^{52}$ erg s$^{-1}$, which
does not explode by the neutrino mechanism, gives an upper limit on the most likely GW emission strength
for massive progenitors that collapse to form black holes without explosion.  Taken at face value, the low convective and SASI
motions of this model imply that, in the absence of rapid rotation,
the GW power emitted by the most massive progenitors that form black
holes could be quite low.

\subsection{GW Energy Spectra and Detection}
\label{section:gwspectra}

\begin{figure}
\epsscale{0.8}
\plotone{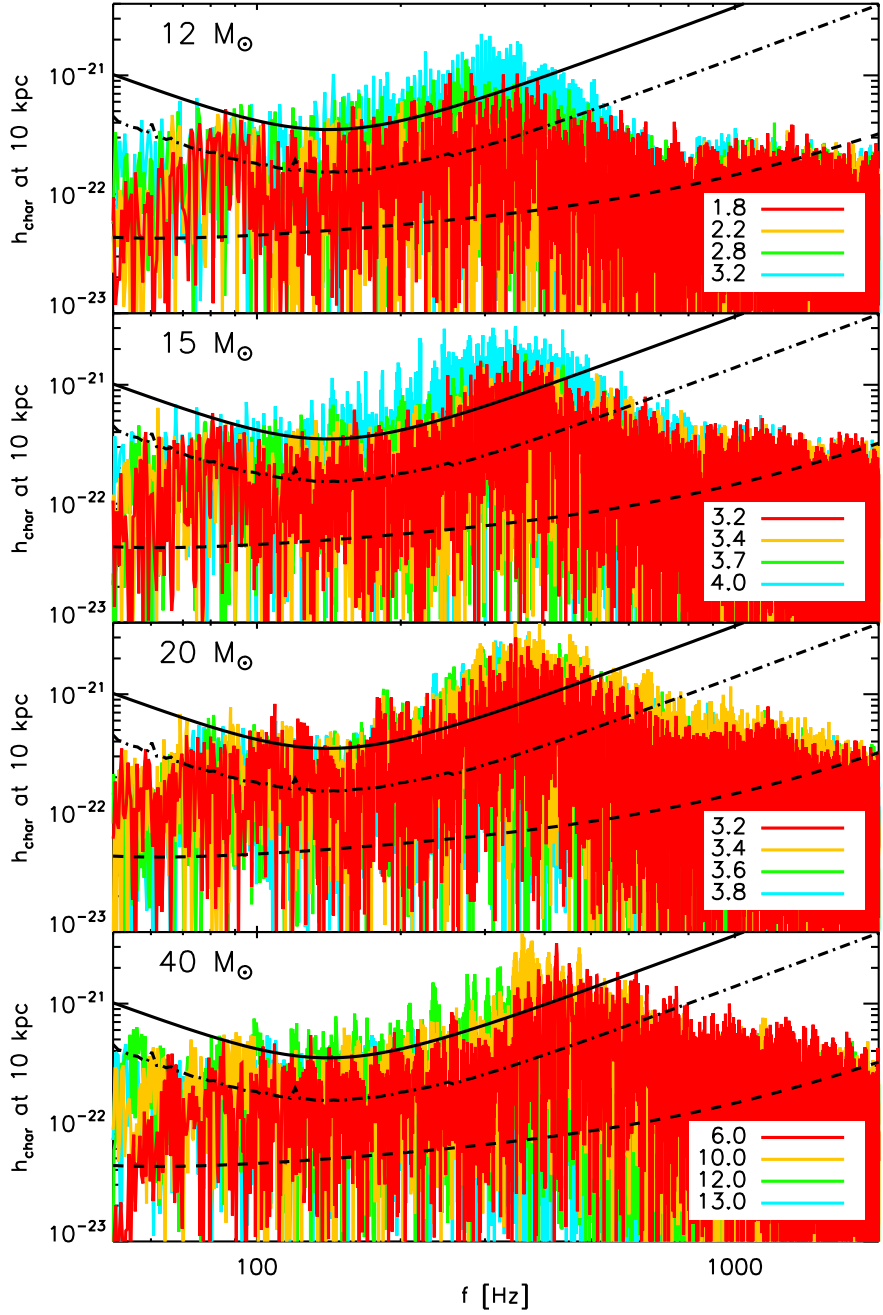}
\caption{$h_{\rm char}$ (eq.~\ref{eq:hchar}) vs. frequency for
  the suite of simulations presented in this paper.  The spectra show
  broad peaks and some dependence upon the progenitor mass: $\sim$300
  Hz for 12 \msun\ and $\sim$400 Hz for 40 \msun.  The power of the
  spectra show only a slight increase (if at all) with neutrino
  luminosity.  For comparison, the approximate noise thresholds for Initial LIGO
  (solid-black curve), enhanced LIGO (dot-dashed-black curve) and
  Advanced LIGO (dashed-black curve) are
  plotted. \label{gwspectra_lumspanel}}
\epsscale{1.0}
\end{figure}

If rotation rates are low and PNS core g-modes are weak, then
postshock-convection/SASI motions are the most probable source of GWs.
In this scenario, the GW emission of Galactic CCSNe ($\sim$10 kpc) is
probably too weak to detect the GW strain waveform directly as a time
series.  Instead, GW observers will search for excess spectral power
above the detector noise in time-frequency maps \citep[e.g., ][]{ligo09}.  
The energy spectra
vs. frequency for all simulations is plotted in
Fig. \ref{gwspectra_lumspanel} assuming a distance to the source, $D$,
of 10 kpc.  Rather than showing $dE_{\mathrm{GW}}/df$, we plot
$h_{\mathrm{char}}$ (eq. \ref{eq:hchar}), which is proportional to
$\sqrt{dE_{\mathrm{GW}}/df}$.  As in Fig.~\ref{hd_lumspanel}, each
panel shows the spectra from simulations for the same progenitor
model, and within each panel the spectra are color-coded and labeled
by $L_{\nu_e}$.  For reference, the noise thresholds for initial LIGO
(solid black line, \citealt{gustafson:99}), Enhanced LIGO
(dot-dashed-black line, \citealt{adhikari:09}), and Advanced LIGO
(burst mode, dashed-black line, \citealt{shoemaker:06}) are included.

\begin{figure}
\plotone{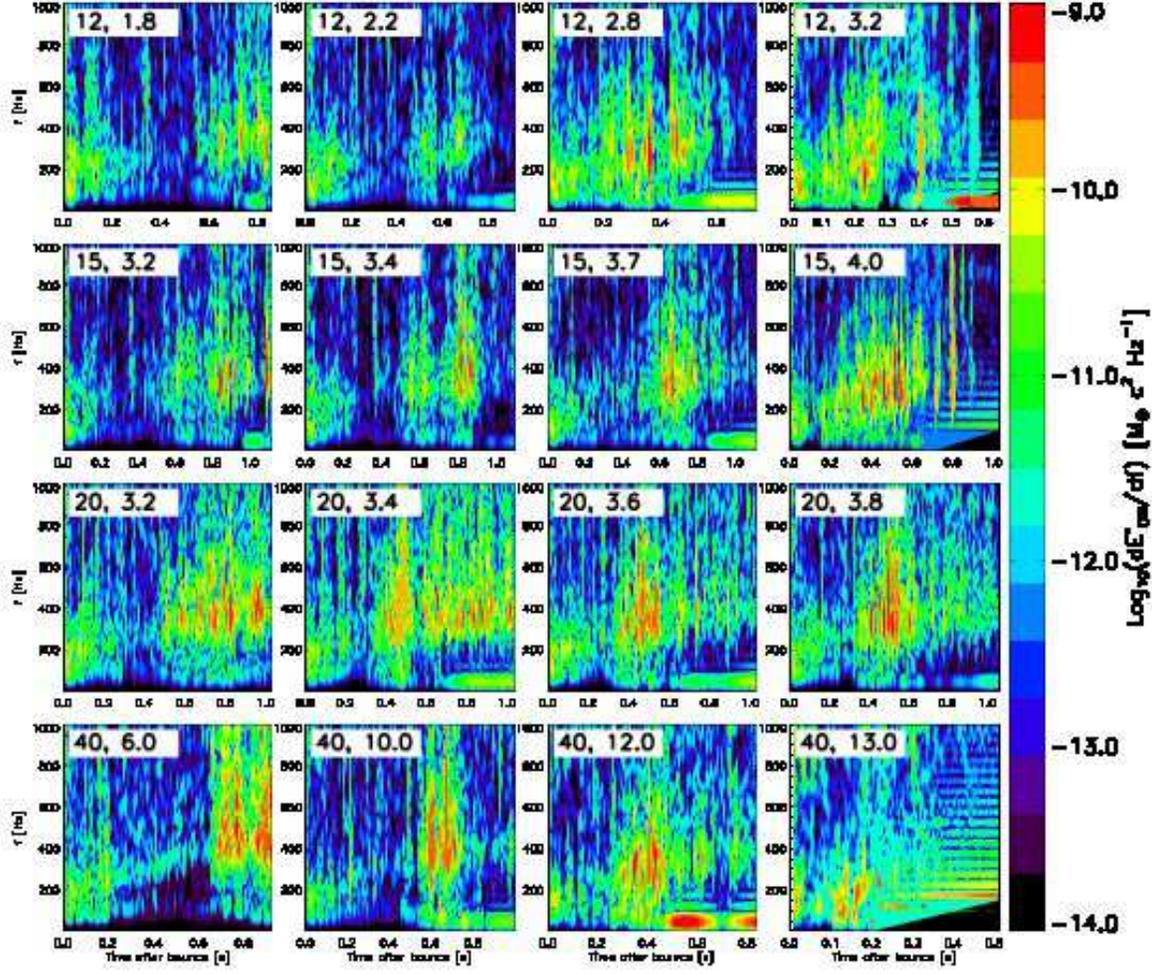}
\caption{Spectrograms of the GW signals showing $dE^*_\mathrm{GW}/df$ as a
  function of frequency (vertical axes) and time (horizontal axes) after
  bounce for the entire set of simulations presented in this paper.
  Bursts of power are associated with prompt convection and the SASI.
  Frequency at peak power increases from $\sim$100 Hz to $\sim$300-400
  Hz, depending upon the explosion time and progenitor used.  We show
  in \S \ref{section:gwsource} that this frequency and the trend to higher frequencies is a consequence
  of the core structure.
  During explosion, GW power transitions to lower frequencies ($\sim$10s
  of Hz).
  \label{gwspectrogram_panel}}
\end{figure}

Characteristically, the spectra show broad peaks with maximum power at
$\sim$300 to $\sim$400 Hz.  The lower and higher frequencies
correspond to lower and higher progenitors masses, respectively.
Later in this section and \S \ref{section:gwsource}, in analyzing the
spectra as a function of post-bounce time, we
show in Fig. \ref{gwspectrogram_panel} that simulations with lower
$L_{\nu_e}$ explode later and obtain higher frequencies at later
times.  However, the spectra in Fig. \ref{gwspectra_lumspanel}, which
are calculated using the entire time sequence, do not clearly show
such a correlation.  The spectra associated with 12-, 15-, and 20-\msun\ progenitors show a second, weaker peak at $\sim$100 Hz that we
associate with prompt convection.  Generally, all spectra are only
marginally above or just below the design noise threshold for Initial
LIGO.  On the other hand, all simulations have power above the design
noise threshold for Enhanced and Advanced LIGOs, with maximum spectral
signal-to-noise ratios of $\sim$5 and $\sim$10-20, respectively.

Characteristic GW amplitudes, frequencies, optimal theoretical
single-detector signal-to-noise ratios, and energies for all
simulations presented in this paper are listed in Table
\ref{table:gwchar}.  The first two columns identify the progenitor
mass and neutrino luminosity of the simulation.  Following are the
maximum amplitude of the GW strain ($|h_{+,\mathrm{rms}}|$), the
signal-to-noise ratios with respect to the Initial, Enhanced, and
Advanced LIGO sensitivity curves (S/N$_\mathrm{LIGO}$,
S/N$_\mathrm{eLIGO}$, and S/N$_\mathrm{advLIGO}$).  We also
provide values for the emitted GW energy ($E_\mathrm{GW}$) and $h_\mathrm{char}$
(eq.~\ref{eq:hchar}) for the maxima
of the characteristic strain spectra and the frequencies $f_\mathrm{char,max}$ at
which they are located.  The values in this table do not present
clean, monotonic trends with progenitor mass or neutrino luminosity.
However, we note some general trends.  More massive progenitors tend
to produce higher frequencies.\footnote{\citet{burrows:07a} show in
Table 1 of their paper a similar trend in that higher progenitor
masses produce higher shock-oscillation frequencies, though at much
smaller values.}  On the other hand, for a given
progenitor model, higher neutrino luminosities give lower
frequencies.  In \S \ref{section:gwsource}, we explain that the
characteristic frequency of GWs is connected with the buoyancy frequency
in the postshock region, which is higher for more massive progenitors,
increases with time, and is largely independent of neutrino
luminosity.  This explains the first correlation with progenitor mass,
but to explain the anti-correlation with neutrino luminosity, we note
that the higher luminosity simulations explode earlier when the
buoyancy frequency is lower.

In general, the typical GW strain for a source at 10 kpc is a few
times $10^{-22}$ to $10^{-21}$, and typical values of $E_\mathrm{GW}$ range
from a few times $10^{-11}$ to a few times $10^{-10}$ \msun\ c$^2$.
\citet{kotake09} conclude that the stochastic properties of the SASI
preclude any relationship between the neutrino luminosity and the GW
amplitude.  While there appears to be no trends, we note several
competing dependencies on neutrino luminosity that result in a
non-monotonic and complicated relationship.
In \S \ref{section:gwsource}, we connect the GW amplitude to the speed
of plumes, $v_p$, at the base of the SASI region.  In addition, we
note that although lower luminosity simulations take longer to
saturate the SASI motions, the saturation value
eventually achieved is similar for all runs.  Since most of the GW
power is emitted during the phase of strongest postshock convection/SASI motions, the total energy
emitted is determined by the duration of this phase,
which starts earlier for the higher neutrino luminosity, but also
ends much earlier at the higher luminosities, so that the total
duration of the stronger SASI phase, and therefore, $E_\mathrm{GW}$ is a
non-monotonic function of neutrino luminosity.

Focusing on the energy spectrum for an
entire GW strain data set is most appropriate when the signal is
regular in both amplitude and frequency.  However,
Figs. \ref{hd_lumspanel} and \ref{hdlabeled} show that the amplitude
and frequency change substantially in these simulations.  In this
case, a time-frequency analysis, or spectrogram, is more appropriate
(see \S \ref{section:gwextract}).  To obtain energy spectra vs. time we
take the Fourier transform of the GW strain convolved with a Hann
window function with width $\sim$20 ms and sample the time domain,
$\tau$ in eq. (\ref{eq:stft}), at intervals of 2
ms.  In Fig.~\ref{gwspectrogram_panel}, we show color maps of
$dE^*_\mathrm{GW}/df$ (eq.~\ref{eq:spectrogram}) vs. frequency and
time.  The warmer colors in Fig.~\ref{gwspectrogram_panel} reflect
higher power, and vice versa for cooler colors.  In general,
simulations that explode later (i.e. with higher
mass progenitors and/or lower neutrino
luminosities) have GW power at higher frequencies.  During explosion,
the frequency at peak power drops to 10s of Hz, owing to asymmetric
mass ejection.

\begin{figure}
\plotone{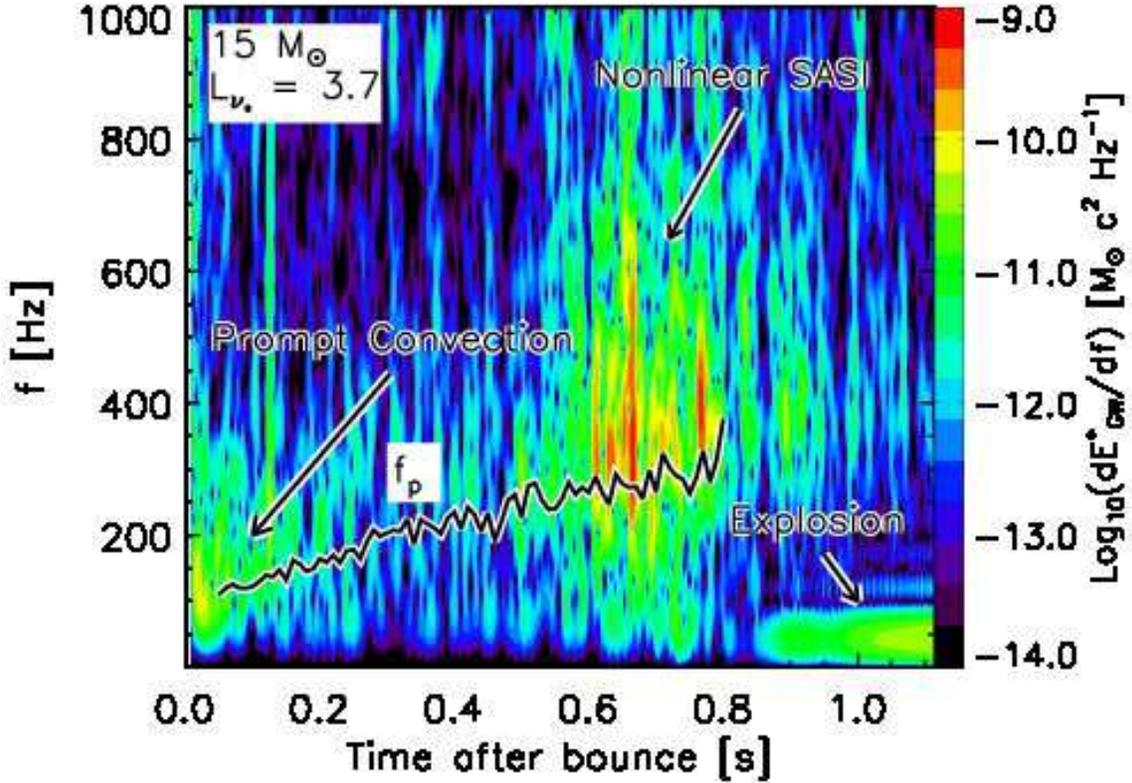}
\caption{Spectrogram, $dE^*_\mathrm{GW}/df$, and characteristic
  plume deceleration frequency, $f_p$, vs.\ time after bounce for the
  simulation using the 15-\msun\ progenitor and $L_{\nu_e} = 3.7
  \times 10^{52}$ erg s$^{-1}$.
  Similar to Fig.~\ref{hdlabeled}, the three features showing the
  largest power are prompt convection just after bounce, nonlinear
  SASI plumes/motions, and explosion.  Unlike in
  the GW strain, the specific morphology of the explosion, other than
  it is asymmetric, is not discernible in the spectrogram.  From
  bounce until explosion, the frequency at peak power agrees with
  $f_p$.  This strong correlation persists whether prompt convection,
  convection, or the SASI are the dominant hydrodynamic processes and
  implies that the plume's buoyant deceleration determines the
  characteristic GW frequency.
\label{gwspectrogram_theory}}
\end{figure}

Unlike in the spectra, which are calculated using the entire time
domain (Fig.~\ref{gwspectra_lumspanel}), prompt convection, nonlinear
SASI, SASI plumes, and explosion are quite apparent in these
time-frequency plots.  Figure~\ref{gwspectrogram_theory} labels these
features in the spectrogram for the simulation using $M = 15$
\msun\ and $L_{\nu_e} = 3.7 \times 10^{52}$ erg s$^{-1}$.  The three
features showing the largest power are prompt convection just after
bounce, nonlinear SASI motions, and explosion.  Though the power
declines between prompt convection and the start of the nonlinear
SASI, the frequency at peak power increases monotonically from
$\sim$100 Hz at bounce to $\sim$300-400 Hz just before explosion.  In
the next section, we propose that the characteristic timescale for the
deceleration of plumes striking the PNS ``surface'' determines the GW
peak frequencies.  The solid black line in
Fig. \ref{gwspectrogram_theory} shows our analytic estimate for this
characteristic frequency, $f_p$, and that it agrees with the frequency
and time evolution of the GW signal from simulation.

\subsection{Source of GWs and a Theory for Their Characteristic Frequencies and Amplitudes}
\label{section:gwsource}

Figure~\ref{hd_totpnssasi} shows that motions associated
with the broad region which includes PNS convection, postshock
convection, and the SASI contribute to the overall GW signal.
\citet{marek09a} come to a similar conclusion.  However, in
\S~\ref{section:signatures} and \S~\ref{section:gwspectra}, we noted that the
strength of the GW signal waxes and wanes as the vigor of postshock
convection and SASI motions rises and falls.  Moreover, spikes in the GW
strain often correlate in time with downdrafts (plumes) striking the
PNS ``surface'' \citep[also noted by][]{marek09a}.  Hence, we argue
that the strength and characteristic frequencies of the GW emission
are determined predominantly by the deceleration of the plumes
striking the PNS ``surface.''

In other contexts, the ``surface'' of the PNS is defined as the
position of the neutrinosphere, the location of a steepening density
gradient, or the narrow region over which the
average radial velocity of the accreting matter approaches zero.
These definitions roughly coincide in space and are
in fact intimately related, but they neither explain nor
quantify the strong decelerations of the downdrafts in an otherwise
continuous medium.  We note, however, that buoyancy forces and their radial profile at the
boundary between convective and stably-stratified regions explain the
deceleration of the plumes and, hence, the GW amplitudes and
frequencies.  We, therefore, define the PNS ``surface'' as the region
where the negatively buoyant downdrafts become positively buoyant and
reverse their downward motions.

An important quantity in analyzing convection and buoyancy is the
square of the Brunt-V\"{a}is\"{a}l\"{a} (or buoyancy) frequency
\begin{equation}
\label{eq:bv}
N^2 = \left ( \frac{G M_r}{r^3}\right ) \left ( \frac{1}{\Gamma_1}
  \frac{d \ln P}{d \ln r} - \frac{d\ln \rho}{d \ln r}\right )\, ,
\end{equation}
where $M_r$ is the mass interior to the radius, $r$, and
the thermodynamic derivative, $\Gamma_1 = (\partial \ln P / \partial
\ln \rho)_S$, is evaluated at constant entropy, $S$.
In the local dispersion relation for
waves, this corresponds to the characteristic frequency
associated with gravity waves (not to be confused with gravitational
waves), whose dominant restoring force is buoyancy,
which sets the frequency scale.  Alternatively, $N^2 < 0$ indicates linear instability and is in
fact the Ledoux (and Rayleigh-Taylor) condition for convection.  Naively,
one might expect the radius where $N^2$ changes sign from negative to
positive to mark the lower boundary of the convection zone.  However,
the Ledoux condition is derived from a linear analysis and ignores
important nonlinear effects.  In practice, convective plumes have
momentum when they reach this boundary and penetrate beyond the Ledoux
condition boundaries.  This is called overshoot and has a long history
and many proposed prescriptions, though many of these are either
inappropriate for the dynamics in core-collapse SNe or lack adequate
physical motivation, resulting in a free parameter (see
\citet{meakin07b}, \citet{arnett09}, and references therein).  For this paper,
we adopt an analysis based upon buoyancy and the bulk Richardson number
\citep{meakin07b}, which is physically motivated and contains no free
parameters.

Beyond the boundaries defined by the Ledoux condition, $N^2$ is
positive and, therefore, coherent plumes that penetrate this boundary
experience a buoyancy force back toward the central regions of
convection.  In this context, the buoyancy acceleration, $b(r)$, felt
by a Lagrangian mass element displaced from $r_i$ to $r$ is related to
the buoyancy frequency by
\begin{equation}
\label{eq:buoyancy}
b(r) = \int^{r_i}_{r} N^2(r) dr \, .
\end{equation}
For plumes approaching the lower boundary of postshock convection, the
magnitude of the integrand is largest where $N^2 > 0$.  Therefore, we
take as $r_i$ the lower Ledoux condition boundary.  With a plume
velocity in the radial direction, $v_p$, at $r_i$ as initial
conditions and eq. (\ref{eq:buoyancy}) as the position dependent
acceleration, we integrate the plume's equation of motion to determine
the depth of penetration, $D_p$.  Because the sources of gravitational
waves are time changing quadrupole motions, the inverse of the
characteristic time of the buoyancy impulse for each plume should correspond roughly to
the peak frequencies in the GW spectra and spectrograms
(Figs. \ref{gwspectra_lumspanel} and \ref{gwspectrogram_panel}).  This
timescale, $t_p$, is defined as the HWHM of the buoyancy acceleration
pulse and the associated frequency is $f_p = 1/(2 \pi
t_p)$\footnote{\label{footnote1}The factor of $2\pi$ arises because
the definition of $t_p$ is similar to the $\sigma_t$ in
$\exp{[-t^2/(2 \sigma_t^2)]}$, and the Fourier transform of this
Gaussian is another Gaussian with width $\sigma_f = 1/(2 \pi
\sigma_t)$.}.  It is this definition of $f_p$ for which we give
quantitative results in this paper.

Using the bulk Richardson number, a dimensionless measure of the
boundary layer stiffness, we obtain an analytic description of $f_p$
that is independent of, and corroborates, our method for calculating
$f_p$.  There is a long tradition of using this dimensionless
parameter in atmospheric sciences to accurately describe the boundary layer in
atmospheric convection \citep{fernando91,fernando97,bretherton99,stevens02}.  More recently
\citet{meakin07b} have successfully used similar arguments to describe
the boundaries of convection in stellar evolution. The bulk
Richardson number is
\begin{equation}
\label{eq:richardson}
R_b = \frac{\Delta b D_p}{v_p^2} \, ,
\end{equation}
where $\Delta b$ is the change in buoyancy acceleration across the
boundary, $D_p$
is a length scale such as the penetration depth, and $v_p^2$ is the
typical velocity of the plume.  This is not to be confused with the
{\it gradient} Richardson number, $R_g$, which is derived from a
linear analysis and is a condition for shear instabilities in a
stratified medium.  While $R_g = 1/4$ is the derived value that
separates stability from instability in a linear analysis, $R_b$ is a
rough measure of the stiffness of a stable layer next to a convective
layer in nonlinear flows.  Approximately, it is the ratio of work per
mass done
by buoyancy ($\Delta b D_p$) and the kinetic energy per mass of the
plume ($v_p^2$).  Therefore, $R_b \lesssim 1$ offers little resistance to
penetration, $R_b > 1$ represents a stiff boundary, and where $R_b
\sim 1$, the work done by buoyancy balances the plume's kinetic energy, causing
it to turn around.  Hence, $R_b$ is
appropriate in characterizing the penetration of downdrafts into the
underlying stable layers.

To estimate scales and trends, we approximate the integral in
eq. (\ref{eq:buoyancy}), giving
\begin{equation}
\label{eq:dbapprox}
\Delta b \sim N^2 D_p \, ,
\end{equation}
and we assume that $R_b \sim 1$, which we argued above is where
buoyancy
balances the plume's kinetic energy and gives the turn-around point.
This gives $(N D_p/v_p)^2 \sim 1$, or
\begin{equation}
\label{eq:dpapprox}
D_p \sim \frac{v_p}{N} \, .
\end{equation}
We note that this scaling is similar to the gravity wave displacement
amplitude derived by \citet{meakin07a}, which they derive by equating
the gravity wave pressure fluctuations to the ram pressure of the
convective eddies at the boundary.  An estimate of the characteristic
timescale is given by the velocity divided by the acceleration, $t_p
\sim v_p/ \Delta b$.  Substituting in eqs. (\ref{eq:dbapprox}) and
(\ref{eq:dpapprox}) gives\footnote{see footnote \ref{footnote1}}
\begin{equation}
\label{eq:fpapprox}
f_p \sim \frac{N}{2 \pi} \, .
\end{equation}
Interestingly, this characteristic
frequency is insensitive to the plume velocity and penetration depth.
Rather, $f_p$ is most sensitive to the buoyancy frequency at the
turning point ($N_{\mathrm{turn}}$).

\begin{figure}
\plotone{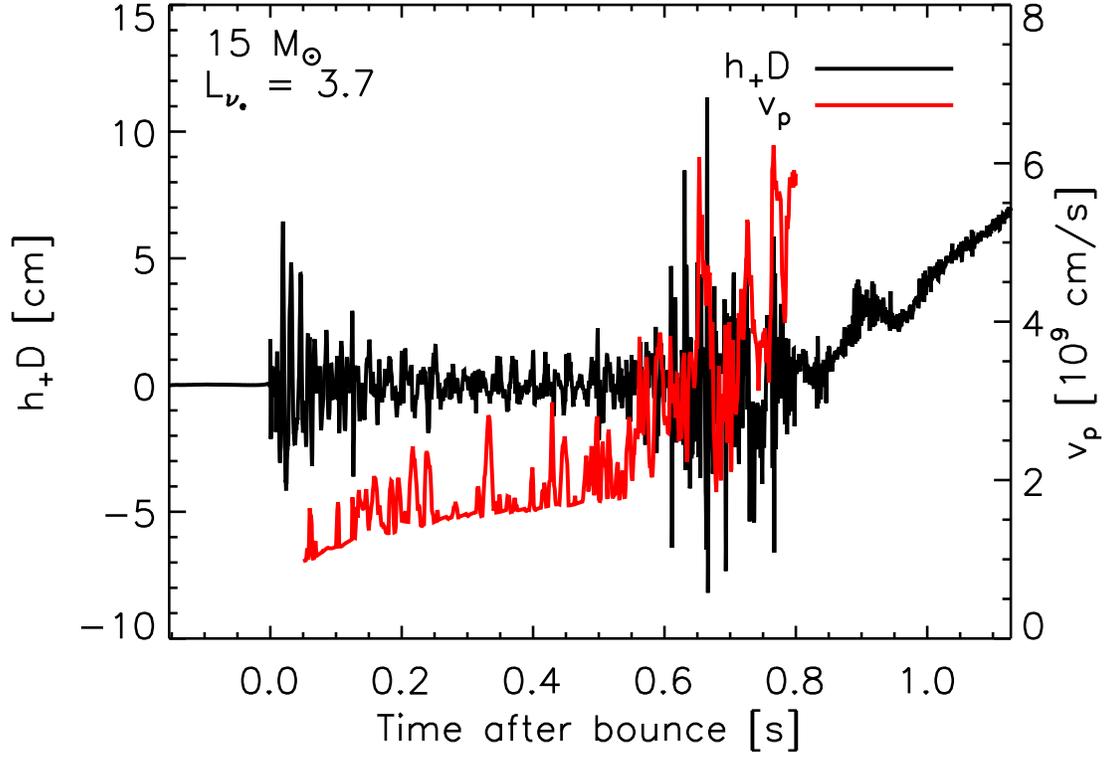}
\caption{Comparison of $h_+D$ with the maximum downward plume speed
below 120 km,
$v_p$, vs. time after bounce.  The
increase in GW wave strain at $\sim$600 ms coincides with the rise in
$v_p$, and many spikes in $h_+D$ coincide with
rapid changes in $v_p$.  A rough estimate of $h_+D$ using these plume
speeds gives $\sim$5 cm, which is consistent with the calculated GW
strain (see the text in \S \ref{section:gwsource}).\label{hdandvpvstime}}
\end{figure}

Although the above analysis suggests that the characteristic frequency
is independent of $v_p$, the amplitude of the GW strain is directly
dependent upon these velocities.  Very roughly, the GW strain is
\begin{equation}
\label{eq:strainapprox}
h_+ \sim \frac{4 \pi G}{Dc^4} \rho \, f_p
\, v_p  \, r^3 \Delta r \, \Delta \mu \, ,
\end{equation}
where we use eqs. (\ref{eq:dibardt}) and (\ref{eq:fmd}), $\Delta \mu$
is the finite difference approximation of $d\cos\theta$, and we assume that
$\partial (\rho v_p) / \partial t \sim \rho \partial v_p / \partial t$
and that $\partial v_p / \partial t \sim f_p v_p$.  Figure
\ref{hdandvpvstime} compares $h_+D$ with the maximum downward plume speed
below 120 km,
$v_p$, as a function of time after bounce.  
The spikes in $v_p$
  correspond to the strongest plumes that strike the PNS ``surface,'' and
  the baseline, below which $v_p$ does not dip, shows the
  time-averaged accretion of material onto the PNS.  As the PNS
  accumulates more mass, exerting stronger gravitational forces, the
  background and plume velocities evolve upward slightly until
  $\sim$550 ms after bounce.  At this time, the motions in the postshock convection/SASI region increase dramatically, leading to a significant rise
  in $v_p$.  After the start of explosion ($\gtrsim$800 ms), it no longer makes
  sense to define a plume velocity, since the GW signal during
  explosion is dominated by explosion dynamics.  Hence, we do not show
  $v_p$ after $\sim$800 ms.

Initially, prompt convection dominates the signal and $v_p$ is
  not relevant.  As prompt convection settles to steady-state
  convection, the plumes become more relevant.  From $\sim$150 ms to
  $\sim$550 ms, both $v_p$ and the GW amplitude are low, but as $v_p$
  grows so does the GW amplitude.  Strikingly, the
increase in GW wave strain at $\sim$600 ms coincides with the rise in
$v_p$.  In fact, many, but not all, spikes in $h_+D$ coincide with
rapid changes in $v_p$.  Furthermore after $\sim$600 ms, substituting values for typical flows near the
boundary (e.g. $\rho \sim 10^{11}$ g
cm$^{-3}$, $v_p \sim 4 \times 10^9$ cm s$^{-1}$, $r \sim 60$ km,
$\Delta r \sim D_p \sim 20$ km, $f_p \sim 300$ Hz, and $\Delta \mu
\sim D_p/(\pi r)$), into eq. (\ref{eq:strainapprox}) gives $h_+D \sim
5$ cm, which is the order of magnitude of the GW strain.

\begin{figure}
\plotone{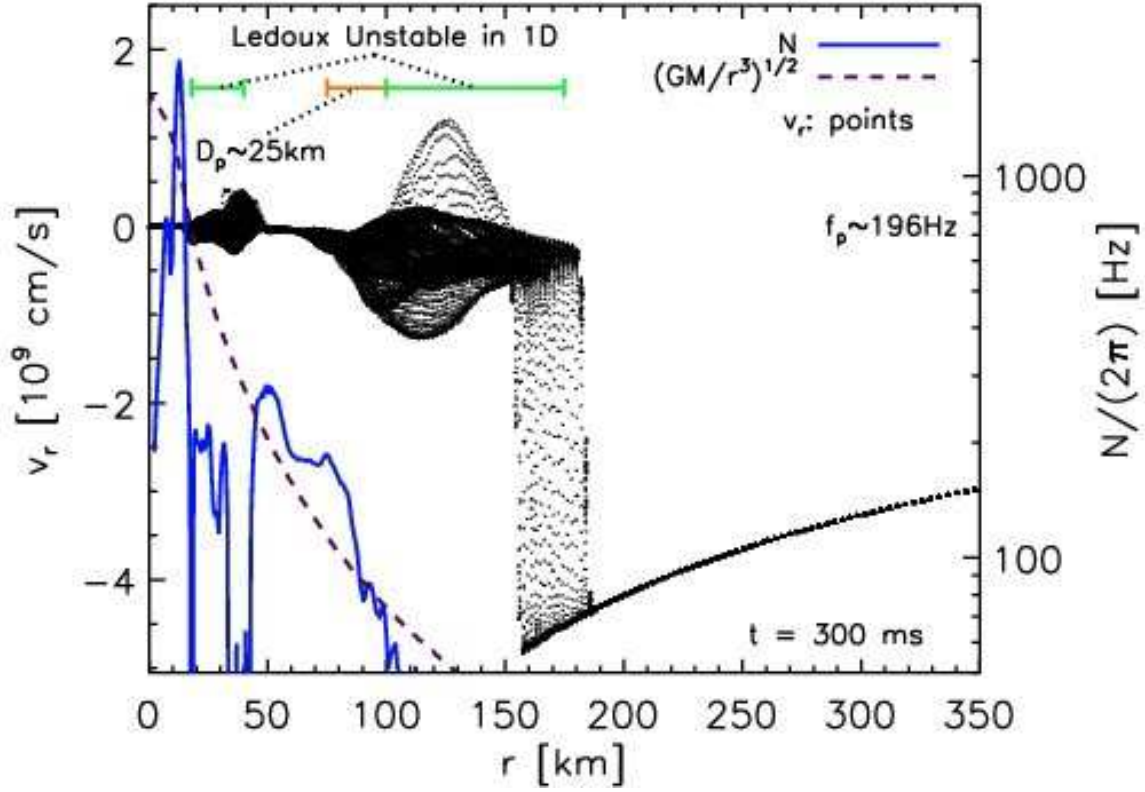}
\caption{The buoyancy frequency, $N$ (blue-solid line, online version), $(GM_r/r^3)^{1/2}$ (purple-dashed
line, online version), and the radial velocity, $v_r$ (black points), vs. radius at 300 ms after bounce for the 2D simulation that uses
the 15-\msun\ progenitor model and $L_{\nu_e} = 3.7 \times 10^{52}$
erg s$^{-1}$.  Starting at larger
radii, the velocity points
show spherical infall, an asymmetric
shock, postshock convection/SASI, a stable layer, the PNS
convection, and a stable inner core.  At
this time the dynamics are dominated by
postshock convection and mild SASI oscillations.
The convection regions are characterized by two regions: 1) that which
is unstable by the Ledoux condition (green lines, online version) and 2) the region of overshoot
below the postshock-convection region (orange line, online version).  
The penetration depth, $D_p$, associated with overshoot
is computed by integrating the plume's equation of motion and finding
the depth at which the plume turns around.  The plume frequency,
$f_p$, obtained from the HWHM timescale of the acceleration pulses, is
$\sim$196 Hz. 
\label{n2velrplot_300}}
\end{figure}

\begin{figure}
\plotone{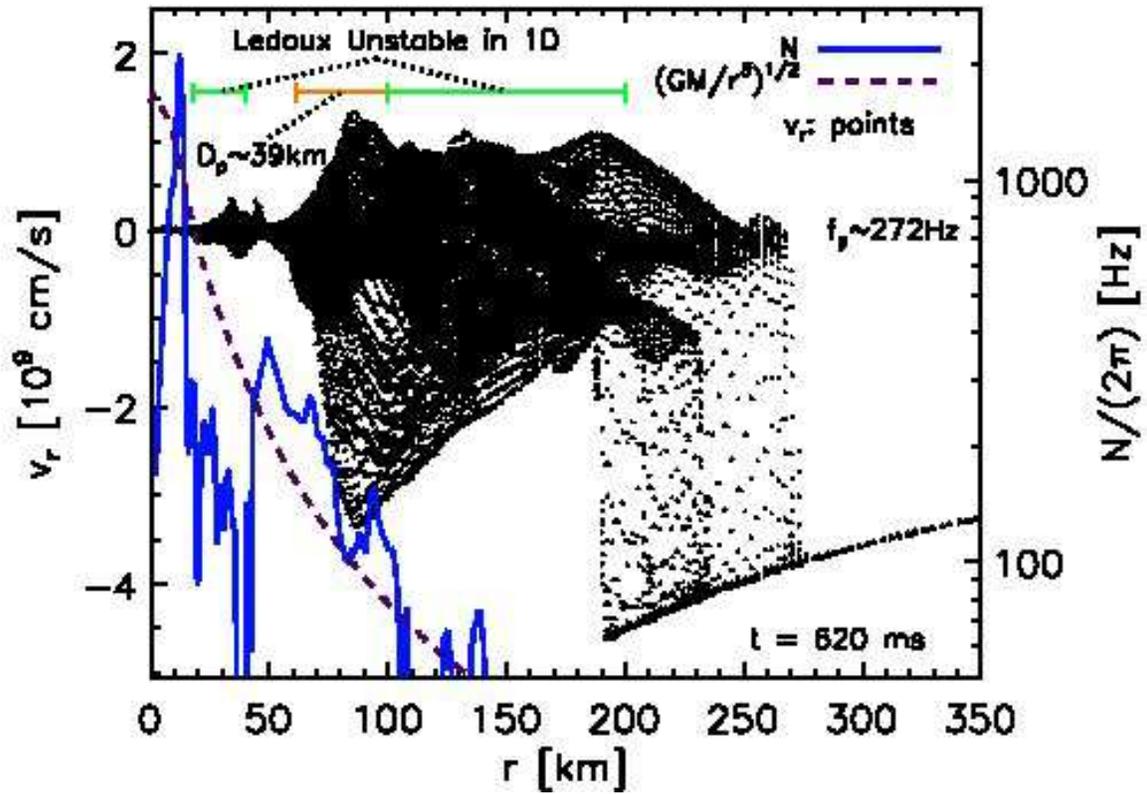}
\caption{Similar to Fig.~\ref{n2velrplot_300}, except that t = 620 ms.
  At this time, the SASI oscillations are much stronger, resulting in a
  higher penetration depth of $D_p \sim 39$ km.  Because $N$ has also
  increased, the characteristic frequency, $f_p$, is higher at 272 Hz.
  \label{n2velrplot_620}}
\end{figure}

Figures \ref{n2velrplot_300} and \ref{n2velrplot_620} depict the
buoyancy frequency, $N$ (solid blue line, online version), $(GM_r/r^3)^{1/2}$ (dashed
purple line, online version), and the radial velocity, $v_r$ (black points), as a
function of radius at 300 and 620 ms after bounce for the 2D
simulation that uses the 15-\msun\ progenitor model and $L_{\nu_e} =
3.7 \times 10^{52}$ erg s$^{-1}$.  The velocities clearly show (from large to small radii),
spherical infall, the asymmetric shock, the postshock-convection/SASI
region, a stable layer, and PNS convection.  At 300 ms
(Fig. \ref{n2velrplot_300}), the dynamics are dominated by postshock
convection and mild SASI oscillations, while at 620 ms, the SASI
oscillations have increased substantially.  Roughly, the buoyancy
frequency scales with the average density,
eq. (\ref{eq:bv}).  However, at the core, the spatial derivatives, and therefore $N$, approach zero.
In the regions of PNS convection ($\sim$20 to $\sim$40 km) and
postshock convection ($\gtrsim$100 km), $N^2$ is negative.  Green line
segments in both figures indicate the regions that are unstable to
convection by the Ledoux condition in 1D simulations, and the
penetration depths, $D_p$, which are computed by integrating the
plume's equation of motion, are labeled and indicated by an orange
line.  At 300 and 620 ms after bounce, $D_p$ at the base of the
region of postshock convection is $\sim$25 and $\sim$39 km, respectively.  These
analytic penetration depths agree with those obtained from the
numerical simulations.  Note that the boundaries of PNS convection
in the 2D simulations coincide with the boundaries given by the Ledoux
condition.  This is due to the fact that
the relatively large buoyancy frequency and small convective speeds
imply very small overshoot depths (eq.~\ref{eq:dpapprox}) at the edges
of PNS convection.  The plume
frequencies, $f_p$, obtained from the HWHM timescale of the
acceleration pulses, are $\sim$196 Hz at 300 ms and $\sim$272 Hz at
620 ms.

\begin{figure}
\plotone{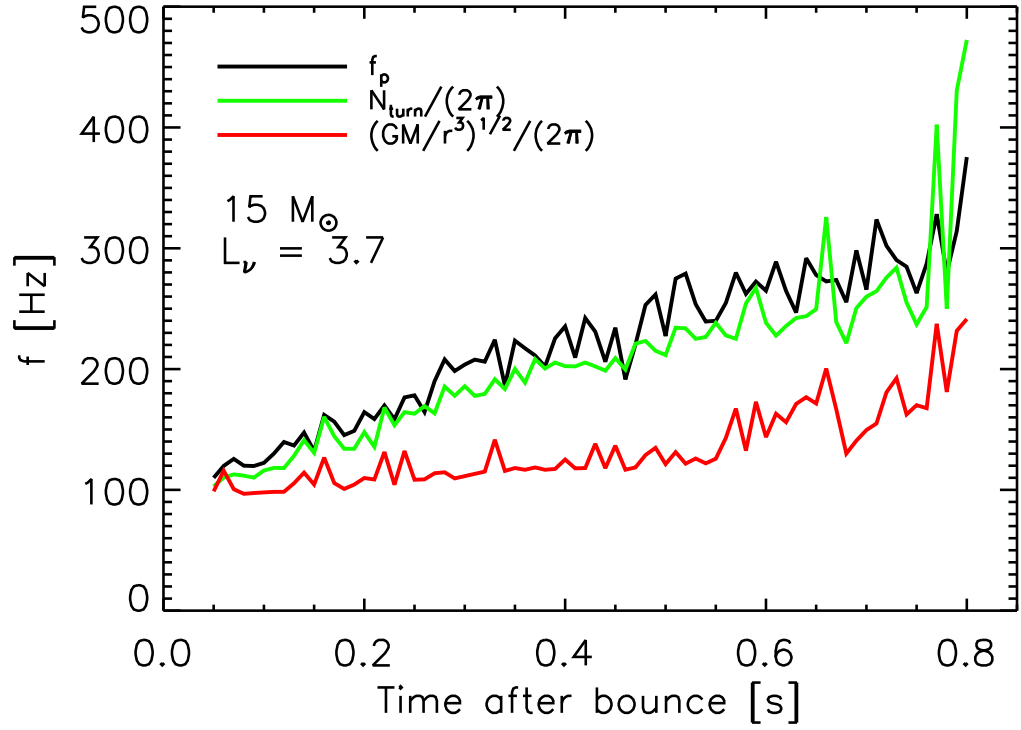}
\caption{Shown are characteristic plume frequency, $f_p$, the buoyancy
frequency at $D_p$, $N_{\rm turn}$, and $(GM_r/r^3)^{1/2}$ at $D_p$ as a
function of time after bounce for the 2D simulation using the 15-\msun\ 
progenitor and $L_{\nu_e} = 3.7 \times 10^{52}$ erg s$^{-1}$.  Initially, all three
coincide.  However, while $N_{\rm turn}$ tracks the time-evolution of
$f_p$, $(GM_r/r^3)^{1/2}$ diverges from the other two measures.  This
indicates that while $(GM_r/r^3)^{1/2}$ (which is related to the
compactness of the PNS) sets the general frequency scale, the local
logarithmic gradients at the penetration depth are just as
important in determining $N_{\rm turn}$.
\label{chartimenturngfreqvstime}}
\end{figure}

\begin{figure}
\plotone{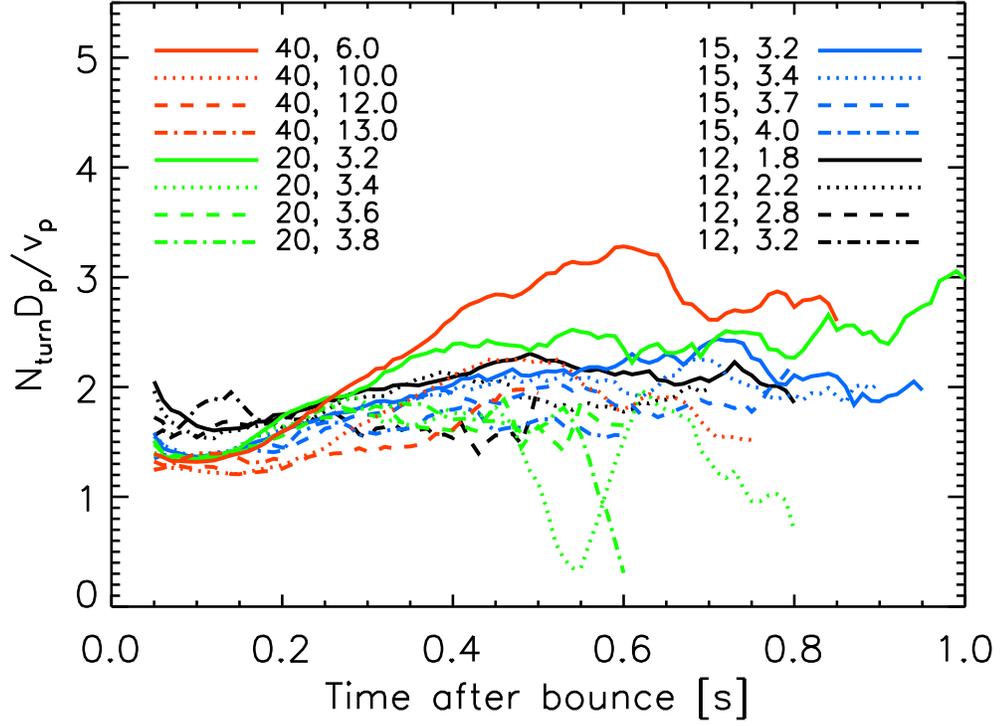}
\caption{The ratio $N_{\rm turn}D_p/v_p$, which is the square
    root of $R_b$ after inserting the approximation that $\Delta b
    \sim N^2 D_p$, vs. time after bounce for all 2D simulations listed
    in Table \ref{table:gwchar}.  Each line is labeled by the
    progenitor mass and neutrino luminosity used.  Each shade (color
    in the online version) represents a set of simulations using the
    same progenitor model: black for 12 \msun, blue (online) for 15
    \msun, green (online) for 20 \msun, and red (online) for 40 \msun.
    The style of line corresponds to the neutrino luminosity used, but
    note that the range of neutrino luminosities for a set of models
    that use the same progenitor mass is different from another set.
    This ratios are $\sim$1.5 to $\sim$2, and change less than 30\%
    during the simulations.  This validates both the assumption that
    $R_b$ is of order 1 where the plumes turn around and the
    approximation that $\Delta b \sim N^2 D_p$.
\label{richardsonvstime}}
\end{figure}

To better understand the contributions to $f_p$, we plot in
Fig.~\ref{chartimenturngfreqvstime} the characteristic frequency
($f_p$), the buoyancy frequency at $D_p$ ($N_{\rm turn}/(2 \pi)$), and
$(GM_r/r^3)^{1/2}/(2 \pi)$ at $D_p$ as a
function of time after bounce.  Strikingly, $f_p$ and $N_{\rm turn}/(2
\pi)$ agree quantitatively.  Since $f_p$ is calculated using the
equation of motion and is independent of the approximations and
assumptions that led to eq. (\ref{eq:fpapprox}), the fact that $f_p$
is nearly equal to $N_{\rm turn}/(2 \pi)$ for all times provides
implicit confirmation of our approximations and assumptions in
deriving the analytic result.  Furthermore, we calculated the ratio
$N_{\rm turn}D_p/v_p$ (the square root of $R_b$ after inserting the
approximation that $\Delta b \sim N^2 D_p$) for all models
(Fig.~\ref{richardsonvstime}), found that
it's value is around 1.5 to 2, and that it changes less than 30\%
during the simulation.  This validates both the assumption that $R_b
\sim 1$ characterizes where the plumes turn around and the
approximation that $\Delta b \sim N^2 D_p$.

Initially, all three measures of frequency in
Fig.~\ref{chartimenturngfreqvstime} coincide.
However, while $N_{\rm turn}/(2 \pi)$ tracks the time-evolution of
$f_p$, $(GM_r/r^3)^{1/2}/(2 \pi)$ diverges from the other two measures.  By plotting both $N_{\rm turn}$ and
$(GM_r/r^3)^{1/2}$, which is the inverse free-fall time, we highlight
global (inverse free-fall time) and local (difference of logarithmic
derivatives) properties of the PNS that contribute to
eq.~(\ref{eq:bv}).  The inverse free-fall time is proportional to the
square root of the average density (compactness), which is strongly
dependent upon the dense-matter EOS.
\citet{marek09a} reported a change in peak GW frequency with stiff and
soft dense-matter equations of state, though they make reference
to only compact PNSs without explaining the timescales in the context of
the buoyancy frequency.  Fig.~\ref{chartimenturngfreqvstime} shows
that both the global compactness and the local gradients at the
penetration depth are important in determining $N_{\rm turn}$.
These local gradients are strongly dependent
upon the details of deleptonization and cooling.
Therefore, the time-evolution of the characteristic frequencies is a
strong function of both the dense-matter EOS in the PNS below the
boundary layer and the local microphysics of the boundary layer
itself.

Figure~\ref{gwspectrogram_theory} compares $f_p$ with the
spectrogram, $dE^*_\mathrm{GW}/df$ (eq.~\ref{eq:spectrogram}), of the simulation using the
15-\msun\ progenitor and $L_{\nu_e} = 3.7 \times 10^{52}$ erg s$^{-1}$.  From bounce until explosion,
the frequency at peak power coincides with $f_p$.  This strong
correlation persists whether prompt convection, convection, or the
SASI are the dominant instabilities and implies that the plume's
buoyant deceleration determines the characteristic GW frequency at all
times before explosion.

\begin{figure}
\plotone{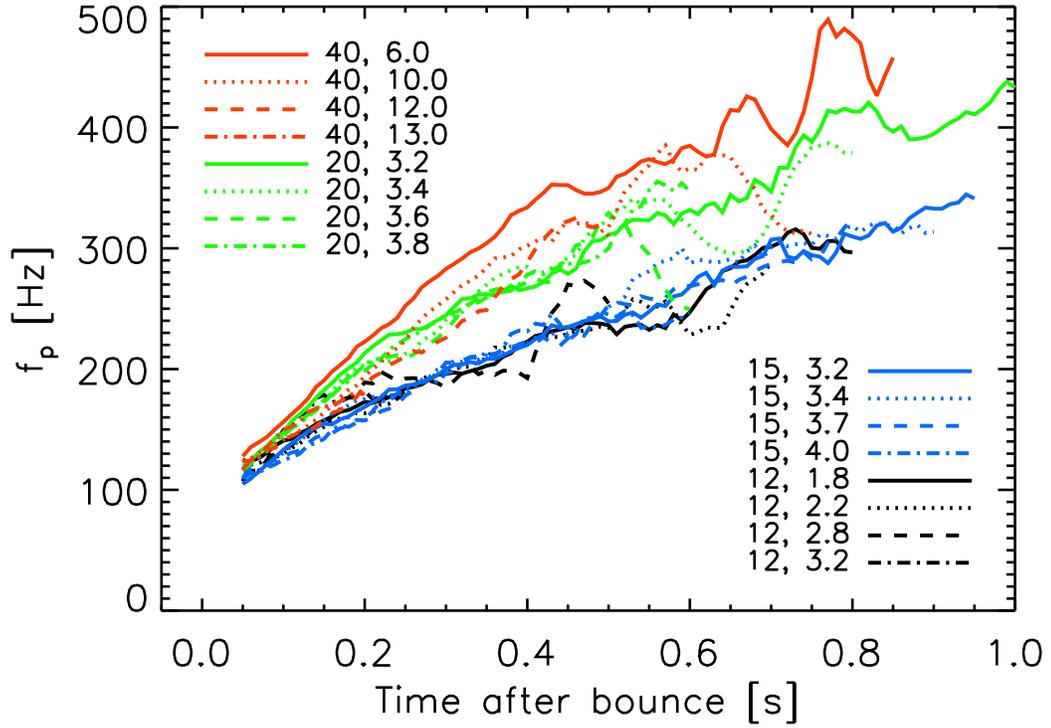}
\caption{Characteristic plume frequency, $f_p$, as a function of time
after bounce for all 2D simulations listed in Table
\ref{table:gwchar}.  The shading (color in the online version) and
line schemes are the same as in Fig.~\ref{richardsonvstime}.
For simulations that use the same
progenitor model, the $f_p$-time evolutions are close.  The
only differences are that the higher luminosity simulations explode
earlier (the end of $f_p$-time curves mark the approximate time of
explosion) and consequently have lower frequencies at explosion.  In
general, the higher mass progenitors produce higher frequencies, with
the exception that the 12- and 15-\msun\ models have very similar
$f_p$-time curves.
\label{fcharvstime}}
\end{figure}

In Fig. \ref{fcharvstime}, we plot $f_p$ as a function of time
after bounce for all 2D simulations.  Each shade (color in the online version) represents a set of
simulations with the same progenitor model
but different $L_{\nu_e}$: black for 12 \msun, blue (online) for 15 \msun, green (online) for 20 \msun, and
red (online) for 40 \msun.  As in Fig. \ref{gwspectrogram_theory}, careful inspection of these $f_p$-time curves show
that they agree with the trends of peak frequency in the GW spectrograms,
Fig. \ref{gwspectrogram_panel}.  For simulations that
use the same progenitor model, their $f_p$ evolutions are practically
indistinguishable.  The only differences are that the higher
luminosity simulations explode earlier (the end of $f_p$-time curves
mark the approximate time of explosion) and consequently have lower
frequencies at explosion.  In general, the higher mass progenitors
produce higher frequencies, with the exception that the 12- and 15-
\msun\ models have very similar $f_p$-time curves.

\citet{yoshida:07} also investigated the interaction between the
plumes of postshock-convection/SASI region and the stable layers below
it.  Their motivation was to investigate the excitation of PNS core
g-modes, which are vital for the success of the acoustic
mechanism \citep{burrows:06,burrows:07b}, by these plumes. 
They concluded that the driving pressure
perturbations had frequencies that are much too low to cause
excitation of the PNS core g-modes.  As a result, they suggest that
core g-mode excitation is inefficient, rather the core g-modes are
forced oscillations.  However, their analysis relied upon the power
spectrum of pressure perturbations at the base of 2D simulations
with inner boundaries at $\sim$50 km.  While they attempt to handle
the subsonic outflow through this inner boundary consistently, they are
forced to set the radial velocity at this inner boundary to maintain
the subsonic structure above the boundary.  This boundary condition
certainly limits the motions at the base of their calculations, and
their unfortunate choice of the location of their inner boundary is
exactly where the plume's motions are most important for exciting core
g-modes.  

Furthermore, there is some confusion in the literature concerning the
appropriate frequencies and excitation mechanism for the core g-modes.
While pressure perturbations play some role in exciting
core g-modes, it is well known that buoyancy is the dominant driving force in
g-waves \citep[e.g.][Ch III, \S 13]{unno89}.  Therefore, velocity
perturbations, which result from both pressure and buoyancy forces,
are more relevant in characterizing the perturbation spectrum at the
surface of the PNS.  The power spectrum of the pressure perturbations
is most useful at highlighting the sounds waves that are emitted by
oscillating PNS.
Furthermore, 
in their discussion of the relevant frequencies, \citet{yoshida:07}
reference the $\sim$30 to $\sim$80 Hz shock-oscillation frequencies in
Table 1 of \citet{burrows:07b} and conclude that motions associated
with these frequencies are much too low to excite core g-modes which
typically have frequencies around 300 Hz.  However, these
shock-oscillation frequencies are not relevant in driving oscillations
at the surface of the PNS.  Rather, we have shown that the
characteristic frequencies associated with the plume impulses are most
relevant and have frequencies that are very similar to the core g-mode
frequencies.  Though we do not address the acoustic mechanism nor the
core g-modes in this paper, in light of our characteristic frequency
analysis, it is possible that the plumes associated with
postshock convection/SASI excite core g-mode oscillations \citep{burrows:07b}.

\section{Discussion and Summary}
\label{section:summary}

In this paper, we present a systematic study of the GW emission from
2D core-collapse supernova simulations in the context of the
convection/SASI-aided neutrino mechanism.  Using a range of progenitor masses
(12, 15, 20, and 40 \msun) and neutrino luminosities, we investigate
the effects of core structure and explosion on the GW signal.  In
addition to GW signatures of prompt convection, PNS and postshock
convection, and the SASI, we characterize the GW signal of asymmetric
mass ejections during explosions and identify the signatures of
explosion that current and near-future GW observatories may detect.
Moreover, we propose that the characteristic GW frequencies and
amplitudes are determined by the deceleration of
postshock-convective/SASI plumes by the buoyantly stable layers below.
We provide gravitational waveforms from all simulated models at
{\tt http://stellarcollapse.org/gwcatalog/murphyetal2009}\,.

For all exploding models, we highlight four distinct phases in the
matter GW strain, $h_+D$.  Figure \ref{hdlabeled} shows that just
after bounce, prompt convection leads to a burst in GW emission with
$h_+D$ reaching $\sim$5 cm and having a frequency of $\sim$100 Hz.
This lasts for $\sim$50 ms or so before steady-state convection is
established and $h_+D$ settles down to $\sim$1 cm.  As the SASI
increases in vigor, $h_+D$ increases to $\sim$10 cm and is punctuated
with strong spikes that are strongly correlated with downdrafts
striking the PNS ``surface.''  From bounce until explosion, the peak
frequency of the GW emission rises from $\sim$100 Hz to $\sim$300-400
Hz, with higher values corresponding to higher mass progenitors and/or
later explosion times (lower neutrino luminosities).

For a few simulations, the transition to strong SASI and GW emission
is initiated by a substantial drop in accretion rate, which is caused
by an abrupt change in density/entropy at the base of the progenitor's
oxygen-burning shell \citep{murphy08b}.  The fact that the correlation
is weak precludes using individual GW detections to investigate the
location of this entropy edge, but a large sample of detections could
provide statistical constraints.  However, if convection/SASI is the
only GW emission mechanism, detection of CCSNe by current and
near-future GW detectors is limited to our galaxy.  Within this
radius, the rate of CCSNe is roughly one per century \citep[e.g.,
][]{vandenbergh91,cappellaro99}, a rate that is far too low to build a
statistical sample.  Such a sample of GW detections from CCSNe is only
feasible if they can be detected with distances out to $\sim$4-5 Mpc,
where the rate increases to $\sim$1-2 per year (e.g.,
\citealt{ando:05}). Third generation observatories, e.g., the proposed
Einstein Telescope\footnote{\tt http://www.et-gw.eu}, or even more
sensitive detectors may be required to study the GW emission from
convection/SASI in CCSNe out to $\sim$4-5~Mpc.

As explosion ensues, the SASI and neutrino-driven convection above the
neutrinosphere subside, causing the rapid oscillations and large
spikes in the GW strain to reduce in amplitude.  In their stead, aspherical mass
ejection results in a slow, but sometimes large, monotonic rise in the
GW strain.  For spherical explosions, the strain settles back to zero
after the onset of explosion, while the strain rises to positive
values for prolate explosions and drops to negative values for oblate
explosions.  This is similar to the ``burst with memory'' GW signal
that asymmetric neutrino emission produces
\citep{braginskii87,burrows96,mueller97}.  Consequently, the peak GW power
emitted moves from $\sim$300-400 Hz to $\sim$10s of Hz.  Initial and
Enhanced LIGO sensitivity at low frequencies inhibit direct detection
of this explosion signature.  Although the improved sensitivity of
Advanced LIGO at lower frequencies might enable direct detection of
the explosion signature, the abrupt reduction in power at the higher
frequencies is the most obvious signature of explosion for the first generation
detectors.

We stress that this reflects the integrated, $\ell = 2$ morphology of the mass
during the explosion.  The snapshots of entropy in
Fig. \ref{prolateoblatestills} show that while the general morphology
of the matter may be prolate, oblate, or spherical, the entropy and
composition distributions are much more complex.  So, while the
shock wave may be spherical, the composition distributions are
likely to be aspherical.  The spherical shock wave but asymmetric Fe
and Si ejecta distributions in the Cas A supernova remnant, might be a
prime example of this \citep{delaney09}.

These monotonic changes due to asymmetric matter ejection will be
superposed on the large secular changes due to asymmetric neutrino
emission \citep{burrows96,mueller97,mueller:04,marek09a}.  If the
timescales and amplitudes are similar, then it will be difficult to
distinguish between the neutrino and asymmetric matter ejection GW
signals.  Simulations that employ more consistent neutrino transport
for several 100 ms past explosion are necessary to resolve this issue.
Regardless, for CCSNe in which convection/SASI is the dominant source
of GWs, distinguishing among prolate, oblate, or spherical
explosions will require direct detection of the GW amplitude to enable
a determination of the sign of the amplitude.  This will only be
possible if a very close post-main-sequence star such as
Betelgeuse, which is a mere $197 \pm 45$ pc away \citep{harper08},
explodes in the near future.
Otherwise, the distances to typical Galactic SNe ($\sim$10 kpc),
detector characteristics, and their sensitivity to
frequencies $\gtrsim$50 Hz imply that the low frequency ``memory''
signal will not be detected, and the signature of explosion will
be the abrupt reduction in GW power.

Except for the explosion signature, \citet{marek09a} report similar
characteristics in the GW strain.  The evolution in frequency from
$\sim$100 Hz just after bounce to larger frequencies at later times
agrees with the results of \citet{marek09a}.  In fact, our higher
frequencies are consistent with the calculations that use a stiff EOS,
but roughly half those reported for their softer EOS simulation.
\citet{kotake07} and \citet{kotake09}, on the other hand, report only
one phase of weak GW emission exhibiting gradual growth and one
characteristic frequency at $\sim$100 Hz.  We suspect that the
idealized initial conditions and omission of the inner 50 km precludes
them from simulating many of the features that we and \citet{marek09a}
identify.

While the total GW emission originates from regions including the
PNS convection at depth, the stable layer above it, and the base of the outer
convective and SASI regions, we find that the characteristic GW
frequencies and amplitudes are, for the most part, established by the
interaction of postshock-convection/SASI region and the stable
layers below.  As plumes in the convective/SASI region fall toward the
stable layer below, momentum carries their motion beyond the lower
boundary of convection as defined by the Ledoux condition.  As they
penetrate the stable regions, the plumes experience a buoyancy
acceleration upward.
It is this impulse that gives rise to
the GW emission.  The pulse width has a
characteristic frequency, $f_p$, that corresponds to the peak power of
GW emission, and we show that $f_p$ is insensitive to the plume
velocity.  Instead, $f_p$ is set by the buoyancy frequency where
the plume reverses its downward motion, $f_p \sim N_{\rm turn}/(2 \pi)$.
Moreover, the conditions during this deceleration
are of the correct order of magnitude to explain the GW strain amplitude.  In
particular, we show that the amplitude is directly proportional to the
plume velocity, $h_+ \propto f_p v_p \propto N_{\rm turn} v_p$.

Since $f_p$ is set by $N_{\rm turn}$, the frequency at peak GW power
gives direct information on the core structure.  This is not too
surprising.  In fact, using soft and stiff dense-matter EOSs,
\citet{marek09a} find higher peak frequencies for the former, which
they attribute to the compactness of the PNS.  
While compactness certainly influences $N_{\rm turn}$, we emphasize that
local conditions at the PNS ``surface'' and their evolution are just
as important in determining the characteristic GW frequencies.
As a point of clarification, we find that sloshing and turnover
timescales do not set the characteristic GW frequency.
Assuming that $f_p$ is proportional
to $v_p$ divided by a length scale such as an eddy size, one might conclude that $f_p$ is
directly proportional to the plume speed.  This contradicts our
earlier finding that $f_p$ is dependent only on the core structure.
However, in \S \ref{section:gwsource}, we
show that the relevant length scale is the deceleration length, or
penetration depth, $D_p$, and that $D_p \sim v_p/N$, giving the result
that $f_p$ is proportional to $N$ and independent of $v_p$.

In general, we find that more massive progenitors have higher $f_p$ at
a given time, though the 12- and 15-\msun\ models show similar $f_p$
evolution.  Interestingly, our simulations show that the $f_p$
vs. time evolution for a given progenitor model is independent of the
neutrino luminosity.  The fact that the characteristic frequency at
explosion is dependent upon neutrino luminosity is simply due to the
fact that
higher luminosity simulations explode earlier in an otherwise similar
$f_p$-time evolution.

Though we state that our conclusions are qualitatively valid, we have made several assumptions that enabled timely calculation of the
simulations in this paper.  The
effects of neutrino transport were approximated by local heating and
cooling functions, azimuthal symmetry was assumed, though nature is
inherently 3D, and we used spherical Newtonian gravity rather than a
full general relativistic treatment.

Differences between 2D and 3D simulations are likely to matter
quantitatively.  For one, in steady-state stellar convection,
 the eddies are larger and
have faster speeds in 2D simulations \cite[and the references therein]{arnett09,meakin07b}.  Although we have demonstrated that the
characteristic frequencies are unlikely to be affected, the GW strain
amplitude is directly proportional to the plume speed and will be
affected.  However, these arguments are for convection alone; they
ignore the fact that in the gain region, convection, the SASI, and
advection are equally important.  For example, the fact that we can
explain the characteristic GW frequencies by considering the
interaction of convective plumes with neighboring buoyant layers
implies that convection is important.  On the other hand, the rapid
rise in plume velocities as the SASI becomes nonlinear is a clear
indication that the SASI is equally important.  Even though
simulations are beginning to explore the 3D aspects of the SASI
\citep{iwakami:08,iwakami:09,kotake09} and advection-modified
convection has been explored in the linear regime \citep{foglizzo:06},
there has yet to be a complete analysis of nonlinear convection that
is modified by either.  Furthermore, \citet{murphy08b} have noted that
the accretion through the convective/gain region imposes an inherent
asymmetry in the convective flow.  While some matter advects quickly
through the convective/gain region (these are the downdrafts/plumes),
some lingers for very long times.  To what degree do these effects
hold in 3D simulations?  As yet the similarities and differences are
not obvious and remain an interesting and important avenue of research
for core-collapse simulations.

Though 3D effects may not alter the characteristic frequencies,
including general relativity and consistent neutrino transport
undoubtedly will.  With general relativistic corrections, the PNS
structure is more compact, leading to higher characteristic
frequencies.  Without proper treatment of neutrino transport, in
particular, $\nu_{\mu}$ and $\nu_{\tau}$ emission, the PNSs in our
simulations do not lose energy and shrink as they should.  Including
these effects should increase $(G M_r / r^3)^{1/2}$ over time, further
increasing $f_p$ in our simulations.
Furthermore, the neutrino transport is important in determining the
structure, and, hence, buoyancy frequency, at the surface of the
PNS. Therefore, to obtain a reliable and quantitative mapping between
the model characteristic frequencies from postshock-convection/SASI
motions and those observed by GW
detectors such as LIGO, approximations to GR and the neutrino
transport that more faithfully mimic nature are required in future
simulations.

\section*{Acknowledgments}

We acknowledge helpful discussions with and input from Casey Meakin
and Eric Agol.  J.W.M. is supported by an NSF Astronomy and
Astrophysics Postdoctoral Fellowship under award AST-0802315.
C.D.O. is supported by a Sherman Fairchild postdoctoral fellowship at
Caltech, by an NSF award under grant number AST-0855535, and by an
Otto Hahn Prize awarded by the Max Planck Society.  A.B. acknowledges
support for this work from the Scientific Discovery through Advanced
Computing (SciDAC) program of the DOE, under grant number
DE-FG02-08ER41544.  We acknowledge that the work reported in this
paper was substantially performed at the TIGRESS high performance
computer center at Princeton University which is jointly supported by
the Princeton Institute for Computational Science and Engineering and
the Princeton University Office of Information Technology and on the
ATHENA cluster at the University of Washington. Further computations
were performed on the NSF Teragrid under grant TG-MCA02N014, on
machines of the Louisiana Optical Network Initiative under grant
LONI\_NUMREL04, and at the National Energy Research Scientific
Computing Center (NERSC), which is supported by the Office of Science
of the US Department of Energy under contract DE-AC03-76SF00098.



\clearpage

\clearpage

\clearpage

\clearpage

\clearpage

\clearpage

\clearpage

\clearpage

\begin{deluxetable}{cccccccccc}
\tabletypesize{\scriptsize}
\tablecaption{
Quantitative Summary of GW Characteristics\tablenotemark{1}
\label{table:gwchar}
}
\tablewidth{0pt}
\tablehead{
  \colhead{Mass\tablenotemark{2}} &
  \colhead{$L_{\nu_e}$\tablenotemark{3}} &
  \colhead{$|h_{+,\mathrm{max}}|$\tablenotemark{4}} &
  \colhead{S/N$_{\mathrm{LIGO}}$\tablenotemark{5}} & 
  \colhead{S/N$_{\mathrm{eLIGO}}$\tablenotemark{6}} & 
  \colhead{S/N$_{\mathrm{advLIGO}}$\tablenotemark{7}} & 
  \colhead{$h_\mathrm{char,max}$\tablenotemark{8}} & 
  \colhead{$f_\mathrm{char,max}$\tablenotemark{9}} &
  \colhead{$E_\mathrm{GW}$\tablenotemark{10}}}
\startdata
12 &  1.8 & 0.27 &  0.8 &  2.0 &  7.9 & 1.04 & 253 &  6.1 \\
12 &  2.2 & 0.20 &  0.7 &  1.7 &  6.7 & 0.61 & 200 &  3.5 \\
12 &  2.8 & 0.46 &  1.2 &  2.7 & 10.8 & 1.35 & 273 &  8.5 \\
12 &  3.2 & 1.13 &  1.6 &  4.0 & 15.7 & 2.23 & 294 & 20.4 \\
15 &  3.2 & 0.37 &  1.1 &  2.6 & 11.4 & 2.14 & 345 & 17.5 \\
15 &  3.4 & 0.40 &  1.0 &  2.4 & 10.2 & 1.49 & 406 & 14.4 \\
15 &  3.7 & 0.37 &  1.1 &  2.8 & 11.5 & 1.83 & 365 & 12.8 \\
15 &  4.0 & 1.53 &  2.1 &  5.3 & 21.2 & 3.10 & 347 & 46.1 \\
20 &  3.2 & 0.32 &  1.4 &  3.4 & 14.5 & 3.00 & 348 & 29.5 \\
20 &  3.4 & 0.70 &  1.8 &  4.5 & 19.5 & 4.68 & 347 & 57.8 \\
20 &  3.6 & 0.56 &  1.5 &  3.9 & 16.1 & 2.67 & 429 & 34.5 \\
20 &  3.8 & 0.48 &  1.5 &  3.8 & 15.7 & 3.14 & 369 & 33.8 \\
40 &  6.0 & 0.33 &  1.0 &  2.6 & 12.2 & 3.23 & 423 & 36.2 \\
40 & 10.0 & 0.68 &  1.4 &  3.6 & 17.3 & 3.93 & 359 & 47.1 \\
40 & 12.0 & 0.82 &  1.4 &  3.4 & 14.0 & 2.04 & 323 & 16.6 \\
40 & 13.0 & 4.53 &  0.8 &  1.8 &  9.4 & 1.08 &   1\tablenotemark{\dag} &  4.1 \\
\enddata
\tablenotetext{1}{This table lists the integrated GW characteristics of the 2D simulations. These
simulations represent a two-dimensional parametrization that
investigates the dependence of GW emission on progenitor mass (column 1) and neutrino luminosity (column 2).}
\tablenotetext{2}{Progenitor model (\msun).}
\tablenotetext{3}{Neutrino Luminosity ($10^{52}$ erg s$^{-1}$).}
\tablenotetext{4}{Maximum GW strain (10$^{-21}$ at 10 kpc).}
\tablenotetext{5}{Optimal theoretical single-detector signal-to-noise using the initial LIGO sensitivity curve \citep{gustafson:99}.}
\tablenotetext{6}{Optimal theoretical signal-to-noise using the Enhanced LIGO sensitivity curve \citep{adhikari:09}.}
\tablenotetext{7}{Optimal theoretical single-detector signal-to-noise using the burst-mode Advanced LIGO sensitivity curve \citep{shoemaker:06}.}
\tablenotetext{8}{Maximum of the characteristic strain spectrum defined in 
eq.~\ref{eq:hchar} (10$^{-21}$ at 10 kpc).}
\tablenotetext{9}{Frequency location of $h_\mathrm{char,max}$ (Hz).}
\tablenotetext{10}{GW energy emitted ($10^{-11}$ \msun\ c$^2$).}
\tablenotetext{\dag}{Because this simulation explodes early and with
  large asymmetry, the low frequency ``memory'' signature in the GW
  strain dominates the energy spectrum.}
\end{deluxetable}

\clearpage

\clearpage

\clearpage

\clearpage

\clearpage

\clearpage

\clearpage

\clearpage

\end{document}